\documentclass[11pt, draftcls, onecolumn]{IEEEtran}
\usepackage{setspace}
\usepackage{graphicx}
\usepackage{xcolor}
\usepackage[normalem]{ulem}

\title{The Asymptotic Behaviour of Information Leakage Metrics}
\author{Sophie~Taylor, Praneeth~Kumar~Vippathalla, and Justin~P.~Coon 
\thanks{The authors are with the Department of Engineering Science, University of Oxford, Parks Road, Oxford, OX1 3PJ, UK, (email: sophie.taylor2@balliol.ox.ac.uk; praneeth.vippathalla@eng.ox.ac.uk; justin.coon@eng.ox.ac.uk).}
\thanks{This research was funded in part by the Engineering and Physical Sciences Research Council under grant number EP/W524311/1, and the U. S. Army Research Laboratory and the U. S. Army Research Office under grant number W911NF-22-1-0070. For the purpose of Open Access, the authors have applied a CC BY public copyright license to any Author Accepted Manuscript (AAM) version arising from this submission.}}
\date{}

\usepackage{cite}
\usepackage{amsmath,amsfonts,amssymb,mathrsfs}
\usepackage{algorithmic}
\usepackage{graphicx}
\usepackage{bm}
\usepackage{amssymb}
\usepackage{threeparttable}
\usepackage{multirow}
\usepackage{mathtools}
\newcommand{\defeq}{\vcentcolon=}
\usepackage{xfrac}
\usepackage{tablefootnote}
\usepackage{comment}
\usepackage{dsfont}
\usepackage{tikz}
\usetikzlibrary{positioning}
\usepackage{subfigure}
\usepackage[english]{babel}
\usepackage{amsthm}
\usepackage[left=3cm, right=3cm, top=3cm]{geometry}
\usepackage{xcite}
\usepackage{enumitem}
\usepackage{array}
\newcolumntype{C}[1]{>{\centering\arraybackslash}p{#1}}
\makeatletter
\newcommand*\bigcdot{\mathpalette\bigcdot@{.5}}
\newcommand*\bigcdot@[2]{\mathbin{\vcenter{\hbox{\scalebox{#2}{$\m@th#1\bullet$}}}}}
\makeatother

\newenvironment{axioms}
 {\enumerate[label=\textbf{A\arabic*.}, ref=A\arabic*]}
 {\endenumerate}
\makeatletter
\newcommand\varitem[1]{\item[\textbf{A\arabic{enumi}\rlap{$#1$}.}]%
  \edef\@currentlabel{A\arabic{enumi}{$#1$}}}
\makeatother
\allowdisplaybreaks

\begin{document}

\maketitle
\begin{abstract}

Information leakage metrics quantify the amount of information about a private random variable $X$ that is leaked through a correlated variable $Y$.
They can be used to evaluate the privacy of a system in which an adversary, from whom $X$ should be kept private, observes $Y$. Global information leakage metrics quantify the overall information leaked upon observing $Y$, whilst their pointwise counterparts define leakage as a function of the \textcolor{black}{particular} realisation \textcolor{black}{$Y=$} $y$, and thus can be viewed as random variables. We consider an adversary who observes many conditionally independent identically distributed realisations of $Y$. We formalise the essential asymptotic behaviour of an information leakage metric, considering in turn what this means for pointwise and global metrics. With these requirements in mind, we take an axiomatic approach to defining a set of pointwise leakage metrics, and a set of global leakage metrics constructed from them. The global set encompasses many known measures including mutual information, Sibson mutual information, Arimoto mutual information, maximal leakage, min entropy leakage, $f$-divergence metrics, and g-leakage. We prove that both sets follow the desired asymptotic behaviour. Finally, we derive composition theorems quantifying the rate of privacy degradation as an adversary is given access to many conditionally independent observations of $Y$. We find that, for pointwise and global metrics, privacy degrades exponentially with increasing observations, at a rate governed by the minimum Chernoff information. This extends the work of Wu et al. (2024), who derived this result for certain known metrics, including some from our global set.

\end{abstract}


\section{Introduction}
Consider a random variable $X$ which contains some sensitive information that should be kept private. Now let $Y$ be a correlated random variable, the value of which will be revealed to a potential adversary. The variables are connected via a noisy channel. To assess the privacy of the system, we must quantify the amount of information about $X$ that is contained in $Y$. There are countless ways in which this can be done, and, as a result, a great deal of privacy or `leakage' metrics have been proposed and studied.

A subset of the measures proposed to evaluate privacy are information theoretic. We define an information theoretic leakage metric as one which quantifies the amount of information about $X$ that is leaked through $Y$. These must be upper bounded by some measure of the information held in $X$, which should depend on the distribution of $X$ only. A simple example of such a metric is the mutual information between $X$ and $Y$, which cannot be larger than the entropy of $X$. Conversely, differential privacy, a metric proposed by Dwork \cite{Dwork}, which has since been extensively studied and even implemented by companies like Apple \cite{apple}, is not an information theoretic leakage measure. It is a ratio of probabilities that can be arbitrarily large given the distribution of $X$. In this work, we consider information theoretic leakage measures.\footnote{We note that \cite{Fernandes} further categorises some metrics as measures of quantitative information flow (QIF). In our work, these are encompassed by the information theoretic leakage group as long as they satisfy the properties outlined. The main example of a QIF measure is g-leakage \cite{Alvim_g,Alvim_g2}.}

Of the proposed privacy metrics, the vast majority are global measures \cite{review80,Issa,Rassouli,Alvim_g,Alvim_g2}. By global, we mean that these measures take the joint distribution of $X$ and $Y$ and return a single number which quantifies the overall privacy of the system. Generally speaking, the information that the adversary learns about $X$ depends on which realisation $y$ they observe. There may be `good' and `bad' $y$'s which reveal little and much about $X$ respectively. It can be beneficial to see this represented in a privacy metric. The solution to this is to view leakage as a random variable. Privacy can be defined in a pointwise fashion, given an observation $Y=y$. As $Y$ is a random variable of which leakage is a function, we obtain a random leakage variable. Rather than a single number, the output contains the full spectrum of possible leakage values along with their probabilities. 
Saeidian \textcolor{black}{et al.}, in \cite{Saeidian_2023}, proposed this approach of viewing leakage as a random variable, and presented pointwise maximal leakage as a possible metric. \textcolor{black}{The authors} explain that this view allows us to make privacy guarantees based on the statistical properties of the leakage distribution, thus allowing for more flexibility than a global measure.

There are myriad ways in which to define privacy; in 2018, \cite{review80} reviewed over 80 proposed metrics. The important point to emphasise is that no single measure can be optimal for all applications. Suppose we have an adversary who will use $Y$ to guess a password $X$. 
Consider as a privacy metric the multiplicative increase in their probability of success upon observing $Y$. 
This metric will look different depending on how many password attempts the adversary is granted. As one can conceive of countless such application examples, we cannot hope to generate a finite list of all reasonable privacy metrics. This motivates an axiomatic view of privacy measures, which is an approach taken by \cite{Issa} in the context of global measures. To the best of our knowledge, such a study on pointwise measures is yet to be done.
We make progress on this topic in this paper.

Successive queries can undermine privacy mechanisms, and attacks are often sequential \cite{wu2020}. An adversary with access to multiple observations of $Y$ may be able to infer more about $X$ than was allowed by design. As an illustration, suppose a data collector has a database containing useful but potentially sensitive health data about a number of individuals. They allow queries to be made on the data by organisations or individuals for research purposes, to which a perturbed response will be provided. An example of a query could be “How many of the individuals are over 65?". The data collector may, for example, add noise to the true answer. Privacy can be degraded if multiple queries are allowed and responses are combined to reduce  uncertainty.
It is therefore of interest to analyse the behaviour of a leakage metric as a result of a number of observations, or in other words, derive composition theorems.
Such results have previously been found for differential privacy \cite{kairouz2015} and for several global information theoretic measures \cite{wu2020}.\footnote{Mutual information, capacity, Sibson mutual information, Arim\textcolor{black}{o}to mutual information, and $\alpha$-maximal leakage.} In this work we study composition theorems for pointwise leakage metrics and for a general class of global leakage metrics.

\vspace{2mm}
\noindent Our main contributions are as follows. 
\begin{enumerate}
    \item We postulate the asymptotic behaviour of a reasonable information leakage metric under the composition of many observations. We consider what this should mean for global and pointwise measures respectively.
    \item Taking an axiomatic view to defining a pointwise information theoretic privacy measure, we outline a set of axioms from which we will prove the desired asymptotic behaviour follows. More generally, the axioms serve as a list of properties that constitute a reasonable pointwise information theoretic privacy measure.
    \item Following our definition of an information theoretic pointwise leakage metric, we define a corresponding set of global metrics. The set includes existing measures such as mutual information, Sibson mutual information, Arimoto mutual information, maximal leakage, min entropy leakage, $f$-divergence metrics, and g-leakage. It retains a great deal of flexibility to include new metrics that may be defined in the future. We prove that all metrics in the set satisfy the desired asymptotic conditions.
    \item We prove that privacy defined according to our pointwise and global definitions degrades exponentially under the composition of many conditionally independent observations for the adversary, at a rate governed by the minimum Chernoff information between distinct conditional channel distributions. This was found to be true of a number of global metrics individually in \cite{wu2020}.
\end{enumerate}
We also draw parallels with the Bayesian approach to hypothesis testing, whereby the probability of error decays exponentially with the Chernoff information between the distribution of each hypothesis.
The adversary's estimate of $X$ can be viewed as a series of hypothesis tests, \textcolor{black}{where $X$ is determined by pointwise comparison, testing all pairings $x$, $x'$.}
Analytic properties \textcolor{black}{regarding hypothesis testing} are well established.

\textcolor{black}{Broadly speaking, the asymptotic behaviour of leakage metrics is practically relevant when an adversary might carry out an inference attack, for example an averaging attack, on privacy, particularly when the number of observations they can access is large. 
Perhaps the most obvious example of such an attack is a side channel attack, in which an adversary gathers information from a system's physical implementation.
Information such as timing, power usage, and electromagnetic field data can form the basis for statistical analysis of the private variable \cite{standaert}. The adversary may take many observations of side channel data to learn something substantive.
Take for example a password $X$ which takes $Y$ seconds to be encrypted. An adversary can use many conditionally independent observations of $Y$ to learn about $X$.
Aside from side channel attacks, other statistical inference attacks also benefit from multiple observations.
Many apps can now collect location trace data upon which statistical analysis could be performed to reveal personal habits. 
This type of sensitive data is more accurately inferred with increasing observations for the adversary.}

\subsection{Notation}
Throughout the paper, random variables will be represented by capital letters and their realisations by lowercase letters.
We denote private variables as $X$ and correlated observed variables as $Y$ with arbitrary but finite alphabets $\mathcal{X}$ and $\mathcal{Y}$ respectively. 
A series of $n$ observed random variables is given by $Y^n \in \mathcal{Y}^n$ where $\mathcal{Y}^n = \mathcal{Y} \times \dots \times \mathcal{Y}$. 
We use $\mathcal{P}_{\mathcal{X}}$ and  $\mathcal{P}$ to represent the set of probability vectors over $\mathcal{X}$ and $\mathcal{Y}$ respectively,
where the subscript $\mathcal{Y}$ is omitted from the latter for brevity as $\mathcal{P}$ appears frequently in proofs.
Defined over $\mathcal{X}$, $E_i$ is a probability vector with $1$ in its $i$th position. This is equivalent to $E_x$ if $x = x_i$.
We use $Q$ with a subscript to represent a true distribution. For example, the joint distribution of $X$ and $Y$ is $Q_{X,Y}$ and the conditional distribution of $Y$ given a realisation $X=x$ is $Q_{Y|X=x}$.

\section{Asymptotic Behaviour of Information Leakage} \label{Section: asymptotic behaviour}
An information leakage metric quantifies the amount of information about a private variable $X$ that an adversary learns through an observed variable $Y$. An adversary cannot learn more than the maximum information contained in the private variable, and as a result information leakage must be upper bounded by a function of the distribution of $X$. We can use these metrics to assess the information leaked about the private variable through a series of observed variables. Let $X \sim Q_{X}$ and $Y \sim Q_{Y}$ where $X \in \mathcal{X}$ and $Y \in \mathcal{Y}$, for finite sets $\mathcal{X}$ and $\mathcal{Y}$.
Suppose that the channel distribution $Q_{Y|X}$ is known, from which $n$ random samples $Y_{1}, Y_{2}, \dots, Y_{n}$ are drawn independent and identically distributed (i.i.d.) according to $Q_{Y|X=x}$. Therefore,
\begin{equation*}
    Q_{Y^{n}|X=x}(y_{1}, \dots y_{n}) = \prod_{i=1}^{n} Q_{Y|X=x}(y_{i}), 
\end{equation*}
where $(y_{1}, \dots y_{n})$ is a realisation of $Y^{n}$. Assume that the adversary observes $Y^n$. The asymptotic behaviour of an information leakage metric refers to its behaviour as $n$ grows large.

An essential property of an information leakage measure is that asymptotically, leakage should not decrease. As $n$ grows large, the adversary becomes more certain of the value of $X$. Asymptotically (i.e., in the limit of large $n$), upon receiving $n$ i.i.d. observations distributed according to $Q_{Y|X=x}$, an adversary knows $X=x$ with as much certainty as is possible through $Y$.\footnote{If two $X$ realisations share a conditional distribution $Q_{Y|X=x}$, the adversary will never distinguish between them through $Y$ alone.} This must not be any less than the certainty with which an adversary knows $X$ with fewer observations. Asymptotically, the information that the adversary learns about $X$ should thus be non-decreasing with $n$. We first discuss what this means for a global measure, before treating the pointwise case. 

Global leakage concerns a private random variable $X$ and a correlated random variable $Y$ whose particular realisation is not specified. 
Global information leakage measures, commonly denoted by $\mathcal{L}(X \to Y)$ generate a single number that quantifies the amount of information about $X$ that is learned through observation of $Y$.
They are real-valued functions defined over the set of all joint distributions of $X$ and $Y$ with finite but arbitrary alphabets. 
Let the global leakage for $n$ i.i.d. realisations of $Y$ be denoted by $\mathcal{L}_{n} \defeq \mathcal{L}(X \to Y^{n})$. It must be that $\mathcal{L}_{n} = g(Q_{X,Y^{n}})$ for some real-valued function $g$. For a fixed joint distribution $Q_{X, Y}$, we require that for integers $n \geq 1$,
\begin{equation} \label{eq: global asymptotic property 1}
    \mathcal{L}_{n} \leq \mathcal{L}_{\infty}
\end{equation}
and
\begin{equation} \label{eq: global asymptotic property 2}
    \mathcal{L}_{n} \to \mathcal{L}_{\infty}
\end{equation}
as $n \to \infty$, where $\mathcal{L}_{\infty}$ is a function of $Q_{X}$ only, and quantifies the amount of information in $X$.
This says that as $n$ grows large, the global leakage about $X$ through $Y^{n}$ approaches the information in $X$ from below.
On top of these asymptotic requirements, we argue that a global leakage measure should satisfy three axioms set out by Issa et al. \cite[p. 1626]{Issa}. These are as follows. Firstly, if $X$ and $Y$ are independent, the leakage through $Y$ about $X$ is zero. We call this the independence axiom. Secondly, if $X-Y-Z$ is a Markov chain, the global leakage from $X$ to $Z$ cannot be greater than that from $X$ to $Y$. This is the data processing axiom. Finally, if $(X_{1}, Y_{1})$ is independent from $(X_{2}, Y_{2})$, then $\mathcal{L}(X_{1}, X_{2} \to Y_{1},Y_{2}) = \mathcal{L}(X_{1} \to Y_{1}) + \mathcal{L}(X_{2} \to Y_{2})$. We call this the additive axiom.
As $X - Y^{n+1} - Y^{n}$ is a Markov chain, it follows from the data processing axiom that a global leakage measure should be monotonic in the sense that $\mathcal{L}_{n} \leq \mathcal{L}_{n+1}$. This says that, overall, leakage cannot decrease by giving the adversary more information. 

Pointwise leakage is defined conditioned on a certain realisation, $Y=y$. The expression $\ell(X \to y)$ quantifies the leakage about $X$ to an adversary who observes $Y=y$. 
It is a function of the joint distribution over $X$ and $Y$ and the particular realisation $Y=y$. 
As $Y$ is a random variable, the pointwise leakage can be treated as one also. 
Let $l(y) \defeq \ell(X \to y)$.
We say that $L \defeq l(Y)$ is the corresponding random leakage variable. Analogously to above, we use $L_{n} \defeq l(Y^{n})$ to denote the random amount of information leaked to an adversary who observes $n$ i.i.d. realisations of $Y$. In the pointwise case, the asymptotic statement is as follows. 
With probability one, there exists some $n' \geq 1$ (which depends on the source realisation) such that for all $n \geq n'$
\begin{equation} \label{eq: pointwise asymptotic property 1}
    L_{n} \leq L_{\infty}
\end{equation}
and
\begin{equation} \label{eq: pointwise asymptotic property 2}
    L_{n} \to L_{\infty}
\end{equation}
as $n \to \infty$ where $L_{\infty}$ is a random variable that is a function of $X$.
This does not appear as strict as the global statement, the reason being simply that pointwise leakage depends on the realisation received. For finite $n$, it is possible that learning $y^{n}$ induces a larger leakage than learning the true value $x$.
To understand this, note that some realisations of $X$ may 
contain more information than others. For instance, it is natural to intuit that more information is gained upon learning $X=x$ if $x$ was initially thought to be very unlikely, than if it was already expected.
Consider the following example, where the true realisation of $X$ is $x_2$. Upon receiving $y^{n}$, an adversary had evidence favouring $X=x_{1}$ with some probability. They then learn with certainty that $X=x_{2}$. If the realisation $x_{1}$ contains more information than $x_{2}$, it is possible that the leakage was higher in the first instance than the second despite the adversary not being certain of $X$.
As we would like to keep pointwise measures as general as possible, we allow those which may add more weight to knowledge gained about realisations that contain more information.
In any case, as $n$ grows, the belief upon receiving $y^{n}$ gets closer to that having observed the true $x$. Eventually, the probability that $X = x_{1}$ becomes insignificant. At this point, $L_{n}$ must be smaller than $L_{\infty}$ as the adversary believes $X=x_{2}$ with high probability, but they are not certain. Intuitively we should see a convergence to $L_{\infty}$.

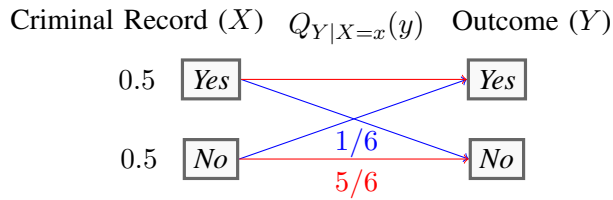
\begin{figure}[ht]
    \centering
        \begin{tikzpicture}[
        squarednode/.style={rectangle, draw=black!60, fill=gray!5, very thick, minimum size=5mm},]
        \node[squarednode] (x1){\textit{Yes}};
        \node[squarednode](x2)[below=0.5cm of x1]{\textit{No}};
        \node[squarednode](y1)[right=3cm of x1]{\textit{Yes}};
        \node[squarednode](y2)[below=0.5cm of y1]{\textit{No}};
        \node(blank)[right=1.4cm of x1]{ };
        \node(QYX)[above=0.2cm of blank]{$Q_{Y|X=x}(y)$};
        \node(p)[left=0.2cm of x1]{$0.5$};
        \node(q)[left=0.2cm of x2]{$0.5$};
        \node(1-s)[red][below=1.4cm of QYX]{$5/6$};
        \node(s)[blue][below=0.85cm of QYX]{$1/6$};
        \node(X)[above=0.2cm of p]{Criminal Record ($X$)};
        \node(Y)[above=0.17cm of y1]{\quad \;\;\;\:  Outcome ($Y$)};

        \draw[->][blue] (x1.east) -- (y2.west);
        \draw[->][blue] (x2.east) -- (y1.west);
        \draw[->][red] (x1.east) -- (y1.west);
        \draw[->][red] (x2.east) -- (y2.west);
        \end{tikzpicture}
    \caption{Joint distribution for criminal record survey response and sanitised output}
    \label{fig: eg criminal record}
\end{figure}

For pointwise measures, we cannot make the statement equivalent to $\mathcal{L}_{n} \leq \mathcal{L}_{n+1}$. To see this more clearly, we illustrate with an example. Suppose a survey is taken by a set of participants, half of which have a criminal record. The survey asks the question “Do you have a criminal record?". So as to preserve plausible deniability, a sanitised version of survey answers is released in which the response to this question is switched with probability $1/6$. The scenario is depicted in Figure \ref{fig: eg criminal record}.
Consider a single participant with no criminal record, and two adversaries. Adversary A is able to access two independent outcomes of the participant's sanitised responses, whilst adversary B can only see the first. In other words, adversary A sees ($y_{1}, y_{2}$) and adversary B sees $y_{1}$. Suppose $(y_{1},y_{2})=$(\textit{No},\textit{Yes}). Adversary B now believes that the participant has no criminal record, with a probability of $5/6$ of her guess being correct. She has learned something about $X$, and intuitively a reasonable pointwise privacy metric will say that the leakage to adversary B is positive. On the other hand, adversary A's belief over $X$ is unchanged. As was the case before making any observations, adversary A cannot favour either realisation of $X$ over the other. It is intuitive to say that he has not learned anything about $X$, and thus that the pointwise leakage should be zero.

\subsection{Global Leakage Example: Mutual Information}
One of the most common and well understood measures of information leakage is mutual information. Mutual information is given by 
\begin{equation*}
    I(X;Y) \defeq \sum_{x \in \mathcal{X}} \sum_{y \in \mathcal{Y}} Q_{X,Y}(x,y) \log \frac{Q_{X,Y}(x,y)}{Q_{X}(x)Q_{Y}(y)}.
\end{equation*}
This is a global information theoretic leakage measure; it takes as input the joint distribution over $X$ and $Y$ and outputs a single number which quantifies the amount of information about $X$ that is learned through $Y$, and it is upper bounded by a quantification of the information contained in $X$. We can say that $\mathcal{L}(X \to Y) = I(X;Y)$. It is established that mutual information satisfies the desired asymptotic properties (\ref{eq: global asymptotic property 1}-\ref{eq: global asymptotic property 2}), as well as monotonicity. We know that $I(X;Y^{n}) \leq H(X)$. Wu et al. \cite[Th. 1]{wu2020} proved that $\mathcal{L}_{n} \to \mathcal{L}_{\infty}$ as $n \to \infty$, where $\mathcal{L}_{\infty} = H(X)$ when the conditional probabilities $Q_{Y|X=x}$ are distinct. Moreover, the convergence occurs at a rate governed by the minimum Chernoff information between distinct conditional distributions $Q_{Y|X=x}$. It is well known that mutual information satisfies the data processing inequality, from which $\mathcal{L}_{n} \leq \mathcal{L}_{n+1}$ follows.

With the help of \cite{wu2020}, the same properties can be verified of several other well established global leakage metrics. In this work, we will prove that they are satisfied for a generalised set of global information theoretic metrics. 

\subsection{Pointwise Leakage Example: Pointwise Maximal Leakage} \label{Section: PL eg PML}
Work on pointwise leakage is far more scarce than that on global leakage in privacy literature. In particular, work regarding the asymptotic properties (\ref{eq: pointwise asymptotic property 1}-\ref{eq: pointwise asymptotic property 2}) does not exist to the best of our knowledge.
Here, we take pointwise maximal leakage (PML) as our example measure. It will be proven in the following section that any pointwise leakage measure defined according to a set of axioms (of which PML is an example) satisfies our required asymptotic properties. In this section we use a numerical example to gain an intuition for how PML behaves under the composition of many \textcolor{black}{conditionally} i.i.d. observations.

PML was introduced as a privacy measure by Saeidian et al. \cite{Saeidian_2023} in 2023, who defined and motivated the metric operationally. The paper also more generally argues the benefits of pointwise measures. Their operational definition is as follows. Let $U$ be some randomised function of $X$. The quantity $\ell_{U}(X \to y)$ is the logarithm of the ratio of the probability of correctly guessing $U$ having observed $y$, to the probability of correctly guessing $U$ with no observations. Pointwise maximal leakage is the supremum of $\ell_{U}(X \to y)$ over all distributions $Q_{U|X}$. Saeidian et al. find that this operational definition is equivalent to the following. With PML as the leakage metric, the amount of leakage of $X$ through a given observation $y$ is
\begin{equation*} 
    l(y) = \log \max_{x: Q_{X}(x)>0} \frac{Q_{X|Y=y}(x)}{Q_{X}(x)},
\end{equation*}
and the associated random leakage variable is $L \defeq l(Y)$. If an adversary is allowed $n$ i.i.d. observations, the leakage becomes
\begin{equation*} 
    l(y^{n}) = \log \max_{x: Q_X (x) > 0} \frac{Q_{X|Y^{n}=y^n}(x)}{Q_{X}(x)}.
\end{equation*}

\begin{figure}[ht]
    \centering
        \begin{tikzpicture}[
        squarednode/.style={rectangle, draw=black!60, fill=gray!5, very thick, minimum size=5mm},]
        \node[squarednode] (x1){$x_{1}$};
        \node[squarednode](x2)[below=0.5cm of x1]{$x_{2}$};
        \node[squarednode](x3)[below=0.5cm of x2]{$x_{3}$};
        \node[squarednode](y1)[right=3cm of x1]{$y_{1}$};
        \node[squarednode](y2)[below=0.5cm of y1]{$y_{2}$};
        \node[squarednode](y3)[below=0.5cm of y2]{$y_{3}$};
        \node(blank)[right=1.4cm of x1]{ };
        \node(QYX)[above=0.2cm of blank]{$Q_{Y|X=x}(y)$};
        \node(X)[above=0.2cm of x1]{$X$};
        \node(Y)[above=0.2cm of y1]{$Y$};
        \node(p)[left=0.2cm of x1]{$p$};
        \node(p)[left=0.2cm of x2]{$q$};
        \node(p)[left=0.2cm of x3]{$r$};
        \node(1-s)[red][below=2.4cm of QYX]{$1-s$};
        \node(s)[blue][below=1.9cm of QYX]{$s/2$};

        \draw[->][blue] (x1.east) -- (y2.west);
        \draw[->][blue] (x1.east) -- (y3.west);
        \draw[->][blue] (x2.east) -- (y1.west);
        \draw[->][blue] (x2.east) -- (y3.west);
        \draw[->][blue] (x3.east) -- (y1.west);
        \draw[->][blue] (x3.east) -- (y2.west);
        \draw[->][red] (x1.east) -- (y1.west);
        \draw[->][red] (x2.east) -- (y2.west);
        \draw[->][red] (x3.east) -- (y3.west);
        \end{tikzpicture}
    \caption{Joint distribution for PML simulation with $(p,q,r)=(0.6,0.3,0.1)$, $s=0.4$}
    \label{fig: setup sim 3x}
\end{figure}
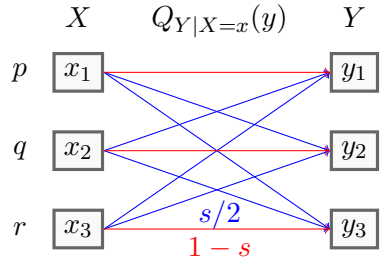

\begin{figure}[ht]
    \centering
    \includegraphics[scale = 0.35]{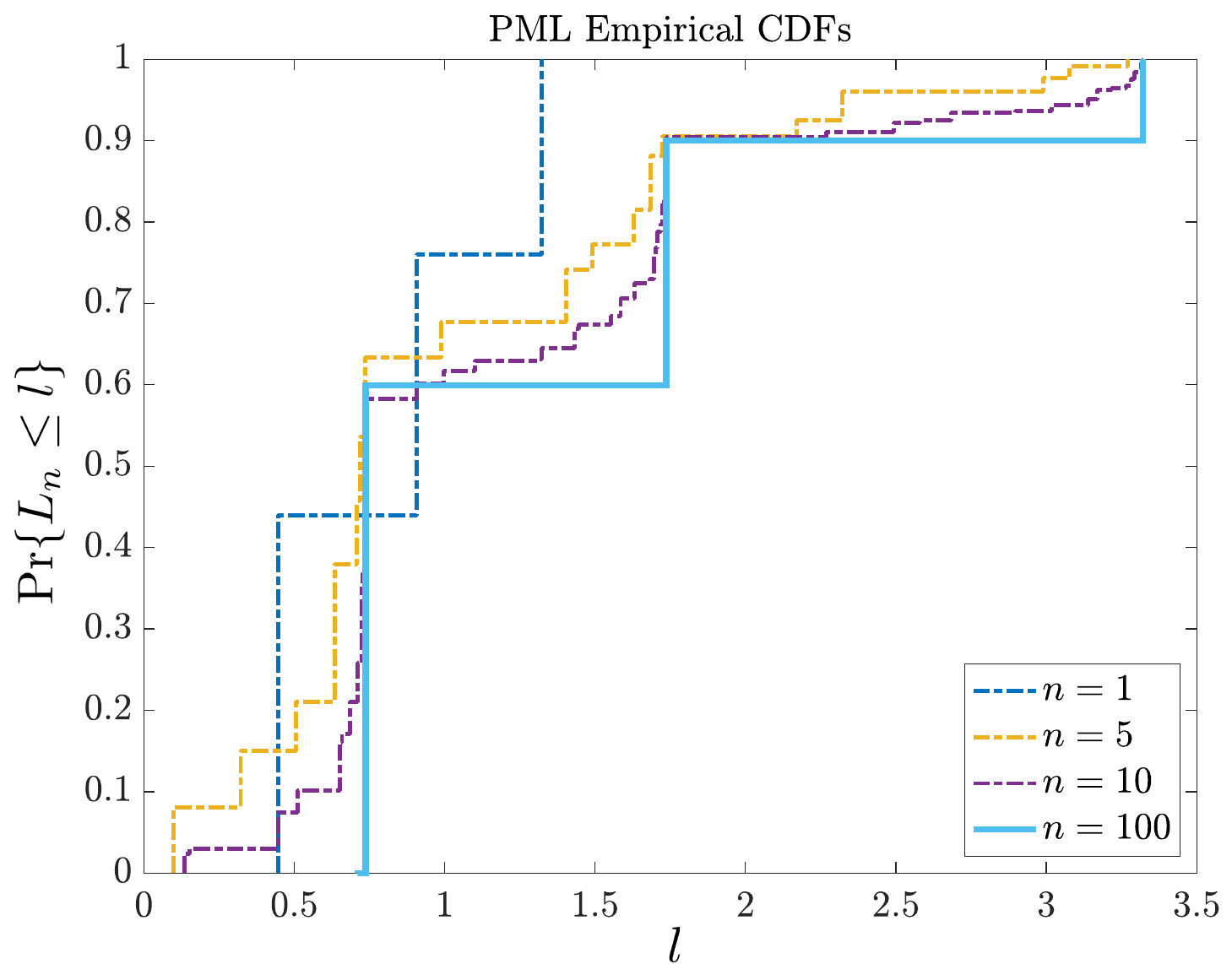}
    \caption{Empirical CDFs of PML as $n$ increases for the joint distribution specified in Fig. \ref{fig: setup sim 3x}}
    \label{fig: PML sim}
\end{figure}

Consider the discrete distribution $Q_{X,Y}$ in Figure \ref{fig: setup sim 3x}. 
Through simulation, the empirical cumulative distribution functions (CDFs) of the random leakage shown in Figure \ref{fig: PML sim} are produced. 
By $n=100$, $L_{n}$ appears visibly to have converged to
\begin{equation*} 
    L_{\infty} = \log \frac{1}{Q_X(X)}.
\end{equation*}
Three leakage values are possible, each corresponding to a realisation of $x$. The leakage corresponding to $X=x_{i}$ is $\log \frac{1}{Q_{X}(x_{i})}$. We can also see that each leakage value occurs with probability equal to the prior probability of the corresponding $x$. For smaller values of $n$, we see that generally the CDFs approach this limiting CDF as $n$ increases. 

Consider the CDF of $L_{1}$. This leakage can also take three values, this time corresponding to each realisation of $y$. As $y_{1}$, $y_{2}$ and $y_{3}$ increase the posterior probability that $X$ is $x_{1}$, $x_{2}$ and $x_{3}$ respectively, $y_{1}$ induces the largest leakage value and $y_{3}$ the least. In fact, observing $Y=y_{1}$ or $Y=y_{2}$ induces a larger leakage than knowing for certain that $X=x_{1}$. Because the probability of observing $y_{3}$ is smaller than the prior probability of $x_{3}$, we see that the CDF of $L_{1}$ passes below the CDF of $L_{\infty}$ for some leakage values. This is an interesting property of pointwise leakage measures; values of $l$ and $n$ may exist such that $\Pr\{L_{n} \leq l\} < \Pr\{L_{\infty} \leq l\}$. Notice also that the minimum possible leakage is larger when $n=1$ than when $n=5$ or $n=10$. This is because conflicting observations, which can reduce the adversary's certainty, are not possible in the case of a single query.

In the following section we will outline a set of axioms for a pointwise leakage measure from which we prove that our desired asymptotic properties follow. It can be verified that PML satisfies the axioms. 

\section{Pointwise Leakage}
In the previous section we presented a set of asymptotic properties that a reasonable pointwise information leakage measure should have. We have also gained an understanding of how one proposed pointwise leakage measure, PML, behaves under the composition of a large number of observations. The goal is now to prove that PML has the desired asymptotic properties, and further to identify what other reasonable pointwise leakage measures share them. To do this, we will take an axiomatic approach. We define a general pointwise leakage function. We further propose and motivate a set of axioms that dictate that the resulting measure is reasonable and natural. It can be verified that PML is in the set of allowed measures. Next, we will prove that all leakage metrics that follow the set of axioms share the desired asymptotic properties.

\vspace{3mm}
\noindent Recall that $X$ is a private random variable and $Y$ is a correlated public random variable with joint distribution $Q_{X,Y}$. We assume that the conditional distributions $Q_{Y|X=x}$,  $x \in \mathcal{X}$ are all distinct (see Remark \ref{remark: conditional x distributions} for the case when this is not satisfied). A pointwise leakage measure $l(y)$ quantifies the resulting amount of leakage about $X$ upon learning $Y=y$. As $Q_{X}$ and $Q_{X|Y=y}$ represent the adversary's knowledge about $X$ before and after receiving $Y=y$ respectively, it is reasonable to quantify leakage with a function of these two probability vectors. So, we consider a pointwise leakage function of the form
\begin{equation}\label{eq: pointwise leakage def}
    l(y) \defeq f(Q_{X|Y=y},Q_{X}),
\end{equation}
for some real-valued function $f$ defined over $\mathcal{P}_{\mathcal{X}} \times {\color{black}\mathcal{P}^{ \circ}_{\mathcal{X}}}$, where $\mathcal{P}_{\mathcal{X}}$ is the set of probability vectors over $\mathcal{X}$ {\color{black} and $\mathcal{P}^{ \circ}_{\mathcal{X}}$ is the interior of the set $\mathcal{P}_{\mathcal{X}}$ containing probability vectors with non-zero components.}\footnote{\color{black}As the posterior probability $Q_{X|Y=y}$ is absolutely continuous with respect to the prior probability $Q_{X}$, i.e., $Q_{X|Y=y}(x)=0$ when $Q_{X}(x)=0$ for any $x \in \mathcal{X}$, the function $f$ does not carry any leakage interpretation when $Q_{X|Y=y}$ is not absolutely continuous with respect to $Q_{X}$. Therefore, we define $f$ only when none of the components of $Q_{X}$ is zero. This does not result in any loss of generality since if $Q_{X|Y=y}(x)=0$ and $Q_{X}(x)=0$ for some $x$, then the corresponding leakage can defined as in \eqref{eq: pointwise leakage def} by excluding $x$ from the alphabet $\mathcal{X}$.} The first argument of the function is the posterior distribution (having observed $Y=y$) whilst the second is the prior.
As $Y \sim Q_{Y}$ is a random variable, we further denote the pointwise leakage random variable $l(Y)$ by $L$. 

\newtheorem{remark}{Remark}
\begin{remark} \label{remark: conditional x distributions}
    When the conditional distributions $Q_{Y|X=x}$,  $x \in \mathcal{X}$ are not all distinct, we define a new random variable $\bar{X}$ which is obtained by grouping $x$ values that have the same conditional distributions. 
    We can then define pointwise leakage as $f(Q_{\bar{X}|Y=y},Q_{\bar{X}})$,  where $f$ is defined over $\mathcal{P}_{\bar{\mathcal{X}}} \times \mathcal{P}_{\bar{\mathcal{X}}}$. As the conditional distributions $Q_{Y|\bar{X}=\bar{x}}$ are distinct by construction, the analyses carried out in the paper remain valid for this case.
    Note that we can only compare the resulting leakage with the information in $\bar{X}$, rather than the information in $X$, which will never be learned in full.
\end{remark}

This formulation retains enough flexibility for $f$ to be defined in a more complex way. For example, it is perfectly possible to define $f$ according to the operational definition of PML, which, as discussed in Section \ref{Section: PL eg PML}, involves a maximisation over the distribution of a random variable $U$.

\subsection{Axioms of the Pointwise Leakage Function} \label{section: axioms pointwise}
In this section we define and motivate a set of properties that a pointwise leakage function should have. Let $P = (p_{1}, p_{2}, \dots, p_{|\mathcal{X}|})$, $Q = (Q_{k}, q_{2}, \dots, q_{|\mathcal{X}|})$ be a pair of probability vectors over $|\mathcal{X}|$. A reasonable pointwise leakage function $f(P,Q)$ satisfies the following axioms for every $Q  \in {\color{black} \mathcal{P}^{\circ}_{X}}$.
\begin{axioms}
  \item \label{item: A1} $f(Q,Q) = 0$.
  \item \label{item: A2} For any $i \in \left\{1, \ldots, |\mathcal{X}|\right\}$ \textcolor{black}{and any $Q \neq E_i$}, $f(E_{i},Q) > 0$, where $E_{i}$ is a probability vector with $1$ as the $i$th element.
  \item \label{item: A3} For any $k$ probability vectors $P_{1}, P_{2}, \dots, P_{k}$ and any coefficients $\lambda_{1}, \lambda_{2}, \dots, \lambda_{k} \geq 0$ such that $\sum_{i=1}^{k} \lambda_{i} = 1$,
  \begin{equation*}
      f\left( \sum_{i=1}^{k} \lambda_{i}P_{i} , Q \right) \leq \max_{i} f(P_{i},Q).
  \end{equation*}
  \item \label{item: A4} If $q_{i} \geq q_{j}$, then
  \begin{equation*}
      f(E_{i},Q) \leq f(E_{j},Q),
  \end{equation*}
  where $E_{i}$ and $E_{j}$ are probability vectors with $1$ as the $i$th and $j$th element respectively.
  \item \label{item: A5} For any $i \in \{1, 2, \dots, |\mathcal{X}|\}$, $f(E_{i},Q)$ is a strict local maximum of the function $f(P,Q)$.
\end{axioms}
If the posterior distribution is identical to the prior, the adversary's knowledge about $X$ is unchanged and the leakage must be zero. On the other hand, if the adversary learns the value of $X$ with certainty, leakage should be positive. This gives Axioms \ref{item: A1} and \ref{item: A2}. 
\textcolor{black}{We do not exclude negative values for $f(P,Q)$ from our analysis. 
Such functions are useful to analyse; many existing global leakage metrics exhibit this property when expressed in a pointwise manner, as can be verified in Section \ref{section: global leakage examples} (Table \ref{table: existing global metrics}). A negative value for $f(P,Q)$ might be interpreted as an adversary finding the posterior, $Q$, more `confusing' than the prior, $P$; perhaps the observation reduced their confidence in their ability to guess $X$.}\textcolor{black}{\footnote{\textcolor{black}{Whether such functions should strictly be considered \textit{information theoretic} leakage measures is up for discussion.
In any case, the functions are useful and could serve as candidates for pointwise privacy metrics.}}}

To understand \ref{item: A3}, we can write
\begin{equation*} 
    Q_{X|Y\in \{y_{1}, \dots, y_{k}\} } = \sum_{i=1}^{k} \frac{Q_{Y=y_{i}}}{\sum_{j=1}^{k} Q_{Y=y_{j}}} Q_{X|Y=y_{i}} = \sum_{i=1}^{k} \lambda_{i} Q_{X|Y=y_{i}},
\end{equation*}
and so the condition becomes
\begin{equation*}
    f(Q_{X|Y\in \{y_{1}, \dots, y_{k}\}},Q) \leq \max_{i}f( Q_{X|Y=y_{i}},Q).
\end{equation*}
The leakage to an adversary that knows that $Y\in \{y_{1}, \dots, y_{k} \}$ cannot be any higher than the worst case.

We motivate \ref{item: A4} by noting that if the posterior distribution is $E_{i}$, the observer knows with certainty that $X=x_{i}$. The leakage therefore represents the information conveyed by the realisation that $X=x_{i}$, which should increase as the prior probability of $x_{i}$ decreases.

\begin{figure}[ht!]
  \centering
\includegraphics[scale=0.85]{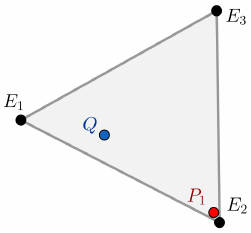}
  \caption{An example over the probability simplex $\mathcal{P}_{\mathcal{X}}$ with $|\mathcal{X}|=3$}
  \label{fig: geogebra axiom simplex}
\end{figure}

Finally, to see \ref{item: A5} we notice that, in a neighbourhood around $E_{i}$, moving closer to $E_{i}$ makes the observer more certain that $X=x_{i}$, which intuitively should increase the information leakage. The region in which this is true must exist but can be arbitrarily small. Consider Figure \ref{fig: geogebra axiom simplex}.
Notice that the position of the prior, $Q$, means that the prior probability of $x_3$, $Q(x_{3})$, is small. 
Intuitively, the $x_3$ realisation contains a lot of information, and 
from many starting positions inside the simplex, moving towards $E_{3}$ should increase the information leakage. 
It is possible that $x_3$ dominates the behaviour of the leakage function throughout the majority of the simplex.
Having said this, however small $Q(x_{3})$ is, it is possible to get close enough to $E_{1}$ or $E_{2}$ such that $P(x_{3}) \ll Q(x_{3})$. At this point, we would not expect $x_{3}$ to dominate the pointwise leakage function. $P_{1}$ is an example of this. In conclusion, in a local region around $E_{i}$ (which may be arbitrarily small), moving $P$ towards $E_{i}$ should strictly increase the pointwise information leakage. This gives us \ref{item: A5}.

The following corollary is a direct consequence of the axioms.

\newtheorem{corollary}{Corollary}
\begin{corollary} \label{corollary: global max}
    \textcolor{black}{For a given $Q\textcolor{black}{\in \mathcal{P}^{\circ}_{\mathcal{X}}}$},
    the global maximum of $f(P,Q)$ is $f(E_{i},Q)$  where $i = \arg \min_{i'}q_{i'}$.
    \begin{proof}
        Any probability vector of length $|\mathcal{X}|$ can be expressed as $\sum_{i=1}^{|\mathcal{X}|}\lambda_{i} E_{i}$
        where $\lambda_{1}, \lambda_{2}, \dots, \lambda_{|\mathcal{X}|} \! \geq 0$, $\sum_{i=1}^{|\mathcal{X}|} \lambda_{i} = 1$. From \ref{item: A3} we know that
        \begin{equation*}
            f\left( \sum_{i=1}^{|\mathcal{X}|}\lambda_{i} E_{i} , Q \right) \leq \max_{i}f(E_{i},Q),
        \end{equation*}
        so the right hand side must be the global maximum. Employing \ref{item: A4} completes the proof. 
    \end{proof}
\end{corollary}
Corollary \ref{corollary: global max} is intuitive. The maximum possible leakage is achieved if we learn with certainty that $X=x_{i}$, where $x_{i}$ was the least likely $X$ realisation according to the prior distribution.

\subsection{Conditions on the Pointwise Leakage Function}

\textcolor{black}{We now consider a class of pointwise leakage functions $f$ that results in exponential privacy decay at a rate governed by the minimum Chernoff information.}
\newtheorem{condition}{Condition}
\begin{condition}\label{assumption: f double bound}
    \textcolor{black}{For each $Q \in \mathcal{P}^{\circ}_{\mathcal{X}}$, there exist neighbourhoods around the extreme points $E_i$, $i \in \{1, \dots, |\mathcal{X}| \}$ of the probability simplex $\mathcal{P}_{\mathcal{X}}$, and constants $A \geq 0$, $B>0$ and $a, b \in \mathbb{R}$ such that the function $f(P, Q)$ satisfies the condition
    \begin{equation}\label{eq: condition 1}
        A(1 - p_i) \left( \log \frac{1}{1- p_i}\right)^{a} \leq f(E_i, Q) - f(P,Q) \leq B(1 - p_i) \left( \log \frac{1}{1- p_i}\right)^{b}
    \end{equation}
    if $P=(p_1, \ldots, p_i, \ldots, p_{|\mathcal{X}|})$ lies in the corresponding neighbourhood of $E_i$.}
\end{condition}

\textcolor{black}{The above condition says that in a neighbourhood around $E_i$, a small change in the first argument of $f$ results in a small change in the output. Note that $1-p_i$ is the maximum difference between any two corresponding elements of $E_i$ and $P$. Furthermore, the constants $A, B, a,$ and $b$ could depend on $Q$. This locally Lipschitz-continuity-like condition is satisfied by most of the examples provided in Section \ref{section: global leakage examples} with appropriate constants, and proving the condition for a given function $f$ tends to be relatively simple. We also note that if $f$ has well-defined derivatives at the extreme points $E_i$ of  $\mathcal{P}_{\mathcal{X}}$, we can directly infer the condition. This is formalised in Proposition~\ref{remark: differentials to conditions}. To state it, we must define what we call the \textbf{derivative property}.\footnote{\color{black} The underlying partial derivatives with respect to the components of the first argument are evaluated in the domain $\mathbb{R}^{|\mathcal{X}|} \times \mathcal{P}^{\circ}_{\mathcal{X}}$, and they exist if the function $f$ has a smooth extension from $\mathcal{P}_{\mathcal{X}} \times \mathcal{P}^{\circ}_{\mathcal{X}}$ to $\mathbb{R}^{|\mathcal{X}|} \times \mathcal{P}^{\circ}_{\mathcal{X}}$. Furthermore, for our purposes, it is enough only to have the partial derivatives defined within $\mathbb{R}_{\geq 0}^{|\mathcal{X}|} \times \mathcal{P}^{\circ}_{\mathcal{X}}$.}}
\newtheorem{definition}{Definition}
\begin{definition}
    The function $f(P,Q)$ satisfies the derivative property if 
for any $Q \in \textcolor{black}{\mathcal{P}^{\circ}_{\mathcal{X}}}$, \textcolor{black}{the function $f(P, Q)$ is differentiable in the first argument at the extreme points $P=E_i$ and }
\begin{equation} \label{eq: derivative propery}
    (U-E_{i}) \cdot \nabla f(E_i,Q) < 0,
\end{equation}
for all $U \textcolor{black}{
\in \mathcal{P}_{\mathcal{X}} }$ such that $U \neq E_{i}$ and $1 \leq i \leq |\mathcal{X}|$. In other words, the directional derivative when evaluated at an extreme point $P=E_i$ in any feasible direction \textcolor{black}{
    within $\mathcal{P}_{\mathcal{X}} $} is negative. Here, $\nabla f(E_i,Q) := \big(\frac{\partial f(P, Q)}{\partial p_1},  \ldots, \frac{\partial f(P, Q)}{\partial p_{|\mathcal{X}|}}\big)\big|_{P=E_{i}}.$
\end{definition}
This turns out often to be the case. Pointwise maximal leakage satisfies the derivative property. \textcolor{black}{The following lemma, whose proof can be found in Section~\ref{section: proof of gradient}, gives an equivalent condition for the derivative property in terms of the partial derivatives at the extreme points.}
\newtheorem{lemma}{Lemma}
\textcolor{black}{
\begin{lemma} \label{lemma: gradients}
    Assume that $f(P, Q)$ is differentiable at $P=E_i$ for all $i \in \{1,\ldots, |\mathcal{X}|\}$ and $Q \in \mathcal{P}^{\circ}_{\mathcal{X}}$. Then, $f$ satisfies the derivative property if and only if for all $j \neq i $,
    \begin{equation*}
        \frac{\partial f(P,Q)}{\partial p_{j}} \bigg|_{P=E_{i}} -  \frac{\partial f(P,Q)}{\partial p_{i}} \bigg|_{P=E_{i}} < 0.
    \end{equation*}
\end{lemma}
}
\textcolor{black}{If the pointwise function is differentiable or satisfies the derivative property, then Condition~\ref{assumption: f double bound} holds. This is formalized in the next proposition.}
\newtheorem{proposition}{Proposition}
\textcolor{black}{
\begin{proposition}\label{remark: differentials to conditions}
    \begin{enumerate}
        \item If the function $f(P,Q)$ is differentiable with respect to $P$ at all the extreme points $P=E_i$,  $i \in \{1, \dots, |\mathcal{X}| \}$ for any $Q \in \mathcal{P}^{\circ}_{\mathcal{X}}$, then $f$ satisfies Condition \ref{assumption: f double bound} with $A=0$, $B>0$, and $b=0$.
        \item If the function $f(P,Q)$ satisfies the derivative property, then $f$      
        satisfies Condition \ref{assumption: f double bound} with $A, B > 0$ and $a=b=0$.
    \end{enumerate}
\end{proposition}
The proof of Proposition~\ref{remark: differentials to conditions} can be found in Section~\ref{section: proof of differentials to conds}.}

\subsection{Asymptotic Behaviour of Pointwise Leakage and Composition Theorems}
In this section we present the effects on pointwise leakage of the composition of many \textcolor{black}{conditionally} i.i.d. observations. 
Given the full system realisation $(x,y_{1},y_{2},\dots,y_{n})$, the pointwise leakage function is
\begin{equation*}
    l(y^{n}) = f(Q_{X|Y^{n}=y^{n}},Q_{X}),
\end{equation*}
and the pointwise leakage random variable for $n$ i.i.d. observations is $L_{n} \defeq l(Y^{n}).$
Let $E_{x}$ be defined such that $E_{x}=E_{i}$, where $i$ is the index of $x$, i.e., $x_{i}=x$. We define an `information function' which quantifies the amount of information contained in a particular realisation of $X$ as
\begin{equation*} 
    i_{X}(x) \defeq f(E_{x},Q_{X}).
\end{equation*}
Like the pointwise leakage function, we may express this as a random variable. 
We define $I_{X}\defeq i_{X}(X)$ as the information random variable, and note that this is equal to $L_{\infty}$. 

We introduce two theorems that characterise the asymptotic behaviour of the leakage random variable, $L_{n}$. 
\newtheorem{theorem}{Theorem}
\begin{theorem} \label{theorem CDF convergence}
\textcolor{black}{If leakage is defined such that $f$ satisfies Condition \ref{assumption: f double bound},}
the sequence $L_{1}, L_{2},\dots$ of pointwise leakage random variables converges almost surely to $I_{X}$ from below, i.e., with probability one,
\begin{equation*}
    L_{n} \to I_{X}
\end{equation*}
as $n \to \infty$, and there exists some $n' \geq 1$
(which depends on the source realisation) such that for all $n \geq n'$,
\begin{equation*}
    L_{n} \leq I_{X}.
\end{equation*}
\end{theorem}
This essentially says that if the adversary is allowed an arbitrarily large number of i.i.d. observations, they will learn all available information\footnote{\textcolor{black}{Recall from Remark \ref{remark: conditional x distributions} that if the conditional distributions $Q_{Y|X=x}$ are not distinct, we can only learn $\bar{X}$. Thus Theorem \ref{theorem CDF convergence} says that $L_n \to I_{\bar{X}}$ and $L_n \leq I_{\bar{X}}$.}} about $X$. 
Notice also that Theorem \ref{theorem CDF convergence} aligns with the desired asymptotic properties for a pointwise information leakage measure outlined in Section \ref{Section: asymptotic behaviour}. Proof of Theorem \ref{theorem CDF convergence} can be found in Section \ref{section: proof of th 1}.

Theorem \ref{theorem: rate of L1 convergence} concerns the rate of convergence of $L_{n}$ to $I_{X}$. This is evaluated by comparing the CDFs.
Let
\begin{align}
    F_{L_{n}}(l) &\defeq \Pr\{L_{n} \leq l\} \label{eq: PML CDF}, \\
    F_{I_{X}}(l) &\defeq \Pr\{I_{X} \leq l\}. \label{eq: information CDF}
\end{align}
These will be referred to as the leakage CDF and information CDF, respectively.
We will use the $L^{1}$-norm, $|| F_{L_{n}} - F_{I_{X}} ||_{1}$ to compare the CDFs. To motivate this, consider the following. Once the CDF of $L_{n}$ is identical to that of $I_{X}$, the adversary is learning all available information about $X$. Theorem \ref{theorem CDF convergence} states that asymptotically, $L_{n}$ approaches $I_{X}$ from below, i.e., privacy does not improve with further  observations. This means that, in the limit of large $n$, privacy worsens the closer the CDF of $L_{n}$ is to that of $I_{X}$. 
As the $L^{1}$-norm is a measure of the difference between two vectors, we can say that for large $n$ the rate of decay of the $L^{1}$ norm between the two CDFs is a natural way to quantify privacy degradation. Before stating Theorem \ref{theorem: rate of L1 convergence}, we must state that the Chernoff information between distributions $P_{1}$ and $P_{2}$ is defined as \cite[p.~387]{Cover2006}:
\begin{equation*}
    \textcolor{black}{\mathscr{C}}(P_{1}||P_{2}) \defeq - \min_{0 \leq \lambda \leq 1} \log \left( \sum_{x} P_{1}^{\lambda}(x) P_{2}^{1-\lambda}(x) \right).
\end{equation*}
Note that base $2$ logarithms are used throughout.

\begin{theorem} \label{theorem: rate of L1 convergence}
\textcolor{black}{If leakage is defined such that $f$ satisfies Condition \ref{assumption: f double bound},}
the rate of decay of the $L^{1}$ norm between the pointwise leakage CDF and the information CDF is exponential and governed by the minimum Chernoff information.
\begin{equation} \label{eq: th 2 general}
        \lim_{n \to \infty} \frac{1}{n} \log || F_{L_{n}} - F_{I_{X}} ||_{1} \leq - \min_{x \neq x'} \textcolor{black}{\mathscr{C}} (Q_{Y|X=x}||Q_{Y|X=x'}). 
\end{equation}
Equality is met if $f$ satisfies \textcolor{black}{Condition \ref{assumption: f double bound} with $A>0$}.
\end{theorem}
The minimum Chernoff information has previously been found to govern the rate of convergence of certain global information theoretic privacy measures \cite{wu2020}. Theorem \ref{theorem: rate of L1 convergence} extends these findings to any pointwise measure that satisfies the axioms \textcolor{black}{and condition} given above. Proof of Theorem \ref{theorem: rate of L1 convergence} can be found in Section \ref{section: proof of th2}. 

\section{Global Leakage} \label{section: global leakage}
A global information leakage metric takes as input the joint distribution $Q_{X,Y}$ and outputs a non-negative real number, which quantifies the overall privacy of the system. 
Here, we define a global leakage function in terms of a pointwise leakage function that is of the form given in (\ref{eq: pointwise leakage def}). Given private data $X$, an adversary who observes $Y$, and a joint distribution $Q_{X,Y}$, the global leakage is
\begin{equation}\label{eq: global leakage def}
    \mathcal{L} \defeq g_{2}\big( \mathbb{E}_{Y}\left[ g_{1}\left( l(Y) \right)
    \right] \big),
\end{equation}
where $g_1$ and $g_2$ are real-valued functions and $l(Y)$ is the pointwise leakage function of the form given in (\ref{eq: pointwise leakage def}). 

The idea behind $\mathbb{E}_Y [g_{1}(\cdot)]$ is that the user can combine the set of leakage values for realisations $y \in \mathcal{Y}$ in the way that best suits their application. For example, they should set $g_{1}$ as the identity function if they want a simple average to be taken over $Y$. The function $g_{2}$ is to suit the user's preference. For example, many authors choose to take a logarithm so that they are working with units that are well understood. The combination of $f$, $g_{1}$ and $g_2$ should be chosen such that the result is a reasonable global information leakage measure. 
It should be upper bounded by some quantification of the information held in $X$. It should also be non-negative, and should satisfy the three axioms set out by \cite[p. 1626]{Issa}.
These are the independence, data processing and additive axioms. Also, we would like the asymptotic requirements set out in Section \ref{Section: asymptotic behaviour} to be met.
We address all properties in this section excluding the non-negative and additive properties, which should be verified independently when defining a global metric in the form of (\ref{eq: global leakage def}).

\subsection{Conditions on the Global Leakage Function}
Next, we outline certain properties that $g_1$ and $g_2$ should have so that the resulting global leakage metric is reasonable and natural. 
First, let the function $h$ be defined as the composition of $g_{1}$ and $f$, i.e.,
\begin{equation}\label{eq: h def}
    h(P,Q) \defeq g_{1}( f(P,Q)).
\end{equation}
The contribution from realisation $y$ to the global leakage should increase with increasing pointwise leakage $l(y)$. Thus, $g_{1}$ should be strictly increasing. It is intuitive that the same is true of $g_{2}$. Note that this means $h$ must have the same local maxima as $f$.
So that global leakage is zero when $X$ and $Y$ are independent, we require that $g_{1}$ and $g_{2}$ are defined such that $g_{2}(g_{1}(0)) = 0$. We will see that for many existing global leakage functions, $g_{2} = g_{1}^{-1}$.
We also assert that $h(P,Q)$ must be convex in $P$. This ensures that the global leakage measure satisfies the data processing inequality, which is stated in Corollary \ref{corollary: global data processing}.
\begin{corollary}\label{corollary: global data processing}\textcolor{black}{Let $X-Y-Z$ be a Markov chain with $X$ being the private variable. If global leakage is defined according to (\ref{eq: global leakage def}) with a pointwise leakage function $f$ and with strictly monotonically increasing functions $g_1$ and $g_2$ such that the function $h$ defined according to (\ref{eq: h def}) is convex in the first argument, then}
    \begin{equation*}
       \mathcal{L}(X \to Z) \leq \mathcal{L}(X \to Y),
    \end{equation*}
    i.e., the global leakage about $X$ through $Z$ is no greater than that through $Y$.
    \begin{proof}
The global leakage about $X$ from $Z$ is 
\begin{align}
    g_{2} \left( \sum_{z \in \mathcal{Z}} Q_{Z}(z) h(Q_{X|Z=z},Q_{X}) \right) &=  g_{2} \left( \sum_{z \in \mathcal{Z}} Q_{Z}(z) h\left(\sum_{y \in \mathcal{Y}} Q_{Y|Z}(y|z) Q_{X|Y=y},Q_{X} \right) \right) \label{eq: global data processing l2}  \\
    &\leq g_{2} \left( \sum_{z \in \mathcal{Z}} Q_{Z}(z) \sum_{y \in \mathcal{Y}} Q_{Y|Z}(y|z) h\left( Q_{X|Y=y},Q_{X} \right) \right) \label{eq: global data processing l3} \\
    &= g_{2} \left( \sum_{y \in \mathcal{Y}} Q_{Y}(y) h(Q_{X|Y=y},Q_{X}) \right), \label{eq: global data processing l4}
\end{align}
where (\ref{eq: global data processing l2}) uses the Markov chain and law of total probability, (\ref{eq: global data processing l3}) applies the convexity of $h$ \textcolor{black}{and the monotonicity of $g_2$}, and (\ref{eq: global data processing l4}) is the global leakage about $X$ from $Y$.
    \end{proof}
\end{corollary}

Given an adversary who receives $n$ observations of $Y$ that are i.i.d. given $X$, the global leakage is
\begin{equation*}
    \mathcal{L}_{n} \defeq g_{2}\big( \mathbb{E}_{Y^{n}}\left[ g_{1}\left( l(Y^{n})\right) \right] \big) = g_{2}\left( \sum_{y^{n} \in \mathcal{Y}^{n}} Q_{Y^{n}}(y^{n}) g_{1}\left( f(Q_{X|Y^{n}=y^{n}},Q_{X}) \right) \right).
\end{equation*}
As $X-Y^{n+1}-Y^{n}$ is a Markov chain, it immediately follows from Corollary \ref{corollary: global data processing} that 
$$\mathcal{L}_{n} \leq \mathcal{L}_{n+1} \text{ and thus that }\mathcal{L}_{n} \leq \mathcal{L}_{\infty}$$ for any finite $n$. 
\textcolor{black}{Next, we restrict our focus to functions $g_1$ and $g_2$ that satisfy the following condition. 
\begin{condition}\label{assumption: g2}
    The function $g$ is strictly increasing and differentiable with $g'>0$ over its domain.
\end{condition}}

\subsection{Composition Theorem for Global Leakage}
The following theorem quantifies $\mathcal{L}_{\infty}$ and the rate at which $\mathcal{L}_{n}$ converges to $\mathcal{L}_{\infty}$.
\begin{theorem} \label{theorem: global rate}
\textcolor{black}{Let global leakage be defined as in \eqref{eq: global leakage def} with a pointwise leakage function $f$ satisfying Condition~\ref{assumption: f double bound} and the functions $g_1$ and $g_2$ satisfying Condition~\ref{assumption: g2} such that $h$, as defined in \eqref{eq: h def}, is convex in its first argument. Then global leakage} approaches its limit exponentially, at a rate governed by the minimum Chernoff information.
    \begin{equation} \label{eq: theorem global}
        \lim_{n \to \infty} \frac{1}{n} \log \left(  \mathcal{L}_{\infty} - \mathcal{L}_{n} \right) \textcolor{black}{\leq} - \min_{x \neq x'} \textcolor{black}{\mathscr{C}} (Q_{Y|X=x}||Q_{Y|X=x'}),
    \end{equation}
    where
    \begin{equation*}
        \mathcal{L}_{\infty} \defeq g_{2}\left( \sum_{x \in 
        \mathcal{X}} Q_{X}(x) g_{1} \left( f(E_{x},Q_{X}) \right)\right).
    \end{equation*}
    \textcolor{black}{Equality is met if $f$ satisfies Condition \ref{assumption: f double bound} with $A>0$.}
\end{theorem}
Theorem \ref{theorem: global rate} says that privacy, when defined according to (\ref{eq: global leakage def}), degrades at a rate governed by the minimum Chernoff information between distinct distributions\footnote{\textcolor{black}{Recall from Remark \ref{remark: conditional x distributions} that if the conditional distributions $Q_{Y|X=x}$ are not distinct, we can only learn $\bar{X}$. Thus, $\mathcal{L}_\infty = g_{2}\left( \sum_{\bar{x} \in 
        \bar{\mathcal{X}}} Q_{\bar{X}}(\bar{x}) g_{1} \left( f(E_{\bar{x}},Q_{\bar{X}}) \right)\right)$.}} $Q_{Y|X=x}$ and $ Q_{Y|X=x'}$. 
This result has been verified individually for mutual information, Sibson and Arimoto mutual information, and maximal leakage by \cite{wu2020}.
The proof of Theorem \ref{theorem: global rate} can be found in Section \ref{section: proof of th 3}.

\subsection{Global Leakage Examples}\label{section: global leakage examples}
We can express many existing global privacy measures in the form of (\ref{eq: global leakage def}). Table \ref{table: existing global metrics} illustrates this for mutual information, Sibson mutual information \cite{Sibson1969,Verdu,Liao}, Arimoto mutual information \cite{Verdu,Liao}, maximal leakage \cite{Issa}, $f$-divergence metrics \cite{Rassouli}, min entropy leakage and g-leakage \cite{Alvim_g,Alvim_g2}.
\begin{table}[ht!]
    \centering
    {\renewcommand{\arraystretch}{2}
    \caption{Existing global leakage metrics expressed in the form (\ref{eq: global leakage def}) and the corresponding pointwise leakage functions.}
    \label{table: existing global metrics}

\begin{tabular}{C{25mm} C{59mm} C{28mm} C{5mm} C{15mm}}
 \hline
 Leakage Metric & Standard Definition & $f(P,Q)$ & $g_{1}(z)$ & $g_{2}(z)$\\ [0.5ex] 
 \hline
 Mutual information & $\sum\limits_{x,y} Q_{X,Y}(x,y)\log \frac{Q_{X,Y}(x,y)}{Q_{X}(x)Q_{Y}(y)}$ & $D(P||Q)$ & $z$ & $z$ \\ 

 Sibson, $\alpha \in (1,\infty)$ & $\frac{\alpha}{\alpha-1}\log\! \sum \limits_{y \in \mathcal{Y}} \left( \! \sum\limits_{x} Q_X(x)Q_{Y|X=x}(y)^{\alpha} \! \! \right)^{\!\frac{1}{\alpha}}$ & $\log \left( \! \sum\limits_{i} q_{i} \left( \frac{p_{i}}{q_{i}}\right)^{\!\alpha} \! \right)
 $ & $2^{\frac{z}{\alpha}}$ & $\frac{\alpha}{\alpha - 1} \log z$ \\

 Maximal leakage & $\log \sum\limits_{y} \max\limits_{x} Q_{Y|X=x}(y)$ & $\log \max\limits_{i} \frac{p_{i}}{q_{i}}$ & $2^{z}$ & $\log{z}$ \\

 Arimoto, $\alpha \in (1,\infty)$ & $\frac{\alpha}{\alpha - 1} \log \sum\limits_{y} \left( \frac{ \sum\limits_{x} Q_{X}(x)^{\alpha} Q_{Y|X=x}(y)^{\alpha}}{\sum\limits_{x} Q_{X}(x)^{\alpha}} \right)^{\frac{1}{\alpha}}$ & $\log \frac{ \sum_{i} p_{i}^{\alpha}}{\sum_{i} q_{i}^{\alpha}}$  & $2^{\frac{z}{\alpha}}$ & $\frac{\alpha}{\alpha - 1} \log z$\\

$f$-divergence & $\sum\limits_{y} Q_{Y}(y) D_{\hat{f}}(Q_{X|Y=y}|Q_{X})$ & $D_{\hat{f}}(P||Q)$ & $z$ & $z$\\

Min entropy leakage & $\log \sum\limits_{y} Q_{Y}(y) \frac{\max\limits_{x} Q_{X|Y=y}(x)}{\max\limits_{x} Q_{X}(x)}$ & $\log \frac{\max\limits_{i} p_{i}}{\max\limits_{i} q_{i}}$ & $2^{z}$ & $\log z$\\

g-leakage & $\log \sum\limits_{y} Q_{Y}(y) \frac{\max\limits_{w \in \mathcal{W}} \sum\limits_{x} Q_{X|Y=y}(x) g(w,x)}{\max\limits_{w \in \mathcal{W}} Q_{X}(x) g(w,x)}$ & $\log \frac{\max\limits_{w} \sum\limits_{i} p_{i} g_{i}(w)}{\max\limits_{w} \sum\limits_{i} q_{i} g_{i}(w)}$ & $2^{z}$ & $\log z$\\
 \hline
\end{tabular}}
\end{table}
Note that Sibson mutual information of order $1$ and $\infty$ correspond to mutual information and maximal leakage respectively. Arimoto mutual information of order $\alpha=1$ also gives standard mutual information. The $f$-divergence $D_{\hat{f}}(P||Q)$ is $\sum_{i} q_{i} \hat{f} \left( \frac{p_{i}}{q_{i}} \right)$.\footnote{We use $\hat{f}$ to avoid confusion with the pointwise leakage function $f$.} For a given global measure, the associated pointwise measure may not be unique. For example, mutual information could equally be represented by $f(P,Q) = \sum_{i} (p_{i} \log p_{i} - q_{i} \log q_{i} )$, i.e., $f(Q_{X|Y=y},Q_{X}) = H(X) - H(X|Y=y)$.

{\color{black}
Theorem \ref{theorem: global rate} applies to a  global leakage metric whose corresponding pointwise function $f$ satisfies Condition~\ref{assumption: f double bound}, and whose $g_1$ and $g_2$ satisfy Condition~\ref{assumption: g2} with $h=g_1\circ f$ being a convex function in its first argument. These conditions can be easily verified for the global measures listed in Table \ref{table: existing global metrics}. It is clear from the table that $g_1$ and $g_2$ are strictly increasing and differentiable with strictly positive derivatives, thereby satisfying Condition~\ref{assumption: g2}. Moreover, the function $h(P, Q)=g_1(f(P, Q))$ is convex for all the metrics. For example, in the cases of mutual information and $f$-divergence, the convexity of $h$ in $P$ follows from the convexity of KL divergence and the  convexity of the function $\hat{f}$, respectively. As $\alpha \geq 1$ in the case of Sibson and Arimoto, we can argue the convexity of $h$ using the triangle inequality for the $\alpha$- norm $||\mathbf{z}||_\alpha= \left(\sum_i z_i^{\alpha} \right)^{\frac{1}{\alpha}}$ when $\alpha>1$. For the rest of the metrics, we can simply use the inequality $\max_i (a_i+b_i) \leq \max_i a_i + \max_i b_i$.

For each global leakage metric, Table \ref{table: global metrics ABab} gives a permissible set of constants $A, B, a ,b$ for each of the pointwise measures $f$ to demonstrate fulfillment of Condition~\ref{assumption: f double bound}. Whether $f$-divergence satisfies the condition depends on the function $\hat{f}$, but it will be the case for many choices. An example that satisfies Condition~\ref{assumption: f double bound} with $a = b =0$, $A=\frac{1}{2}$, and $B = \frac{|\mathcal{X}|}{2}$ is $\hat{f}(z) = \frac{1}{2}|z-1|$ which results in total variation distance as a privacy metric \cite{Rassouli}. If $\hat{f}$ is differentiable over its domain, we can apply Proposition~\ref{remark: conditional x distributions} to say that $A=0$, $B>0$ and $b=0$.

\begin{table}[ht!]
    \centering
    {\renewcommand{\arraystretch}{2}
    \caption{Permissible constants $A,B,a,$ and $b$ for the pointwise leakage functions $f$  of existing global leakage metrics to satisfy Condition \ref{assumption: f double bound}.}
    \label{table: global metrics ABab}
\begin{threeparttable}
\begin{tabular}{c c c c c}
 \hline
 Leakage Metric & $A$ & $B$ & $a$ & $b$\\ [0.5ex] 
 \hline
 Mutual information$^a$ & $M$& $2(| \mathcal{X}| -1)$ & $0$ & $1$ \\ 

 Sibson, $\alpha \in (1,\infty)$ & $\frac{\alpha}{2 \ln 2}$ & $\frac{2 \alpha}{\ln 2}
 $ & $0$ & $0$ \\

 Maximal leakage & $\frac{1}{ \ln 2}$ & $\frac{2}{\ln 2}$ & $0$ & $0$ \\

 Arimoto, $\alpha \in (1,\infty)$ & $\frac{\alpha}{2 \ln 2}$ & $\frac{2 \alpha}{\ln 2}
 $ & $0$ & $0$ \\

Min entropy leakage & $\frac{1}{\ln 2}$ & $\frac{2}{\ln 2}$ & $0$ & $0$\\

g-leakage & 
$0$& $2$ & $0$ & $0$\\
 \hline
\end{tabular}
 \begin{tablenotes}[para,flushleft]
  \footnotesize{
    $^a$ $M$ is a function of $Q_X$. 
  }
  \end{tablenotes}
  \end{threeparttable}}
\end{table}

 We provide a couple of proof examples. First, consider maximal leakage, whose pointwise leakage function $f(P,Q)$ is $\log \max\limits_{i} \frac{p_{i}}{q_{i}}$. Thus, the output at the extreme point $E_i$ is $f(E_i, Q) = -\log q_i$.
If $P$ is in a small neighbourhood of $E_i$, where $\left\lVert P-E_i \right\rVert_1=  2 (1-p_i)$ is sufficiently close to zero, we have $f(P,Q) = \log\frac{p_i}{q_i} = f(E_i, Q) + \log(1-(1-p_i)).$
Thus,
\begin{align*}
    \frac{1}{\ln 2}(1-p_i) \leq f(E_i,Q) - f(P,Q) \leq  \frac{2}{\ln 2}(1-p_i).
\end{align*}
Hence, Condition \ref{assumption: f double bound} is satisfied with $a = b = 0$, $A=\frac{1}{\ln 2}$, and $B=\frac{2}{\ln 2}$. Next, we consider mutual information, whose pointwise leakage function is KL divergence. The output at the extreme point $E_i$ is $f(E_i,Q)= -\log q_i$. Starting with the lower bound, as $P$ is in a small neighbourhood of $E_i$, where $p_i$ is sufficiently close to $1$, we have
$
    f(P,Q) \leq -\log q_i + \sum_{j \neq i} p_j \log \frac{p_j}{q_j}.
$
Next note that, as each $p_j$ where $j\neq i$ is small enough, we can bound the terms in the summation as $-p_j \log \frac{p_j}{q_j} \geq M p_j$ for all $j\neq i$ using a constant $M$. Thus,
\begin{align*}
    f(E_i,Q) - f(P,Q) \geq M \sum_{j \neq i} p_j \geq M (1-p_i). 
\end{align*}
Turning our attention to the upper bound, we can write 
$
    f(P, Q) \geq p_i \log \frac{p_i}{q_i} + (|\mathcal{X}|-1) (1-p_i) \log (1-p_i),
$
where we have used the fact that $z\log \frac{z}{q}$ is monotonically decreasing when $z$ is small enough and that for $j\neq i$, $p_j \leq 1-p_i,$ which is close to zero.
Thus,
\begin{align*}
    f(E_i,Q) - f(P,Q) \leq (1-p_i) \log \frac{1}{q_i} + p_i\log \frac{1}{p_i} + (|\mathcal{X}|-1) (1-p_i) \log \frac{1}{1-p_i}.
\end{align*}
The final term is the leading order term. Therefore, Condition \ref{assumption: f double bound} is satisfied with $a=0$, $b=1$, $A = M$ and $B = 2(|\mathcal{X}|-1)$. The remaining pointwise measures can be addressed in a similar way.

Consequently, Table \ref{table: global metrics ABab} confirms that under the composition of many conditionally i.i.d. observations, mutual information, Sibson mutual information, maximal leakage and Arimoto mutual information approach their limits exponentially at a rate equal to the minimum Chernoff information. On the other hand, we can say that for g-leakage, the rate is at least the minimum Chernoff information. 
The rate for a given $f$-divergence function could be verified  through identification of $A, B, a$, and $b$. 
This is generally easy to do, as demonstrated by the proofs above for maximal leakage and mutual information.

It is also worth highlighting that Theorems \ref{theorem CDF convergence} and \ref{theorem: rate of L1 convergence} apply to pointwise leakage functions $f$ subject to their fulfilment of Condition \ref{assumption: f double bound}. Importantly, PML, which corresponds the to maximal leakage metric, satisfies the condition with $A, B>0$ and $a=b=0$. Hence, the rate of decay of the $L^{1}$ norm between the pointwise leakage CDF and the information CDF, given in Theorem~\ref{theorem: rate of L1 convergence} is exactly the minimum Chernoff information.
If the other functions $f$ of Table \ref{table: existing global metrics} were to be used independently as pointwise privacy metrics, their composition properties could be concluded from our results. 
}

\section{Privacy and Hypothesis Testing}
The composition results we have found are reminiscent of similar known results for Bayesian hypothesis testing. Indeed, we can use the latter to develop an intuition for these newer findings. 

Consider a series of variables $Z_{1}, Z_{2}, \dots, Z_{n}$ that are distributed i.i.d. according to $P$. We wish to select one of two hypothesis: $H_{1}: P = P_{1}$ and $H_{2}: P = P_{2}$. We select $H_{1}$ if $z^{n}$ is in the acceptance region, $A_{n} \subseteq \mathcal{Z}^{n}$.
If the hypothesis test is Bayesian, the error probability is \cite[p. 395]{Cover2006}
\begin{equation*}
    P_{e}^{(n)} = \pi_1 P_{1}^{n}(A_{n}^{\mathrm{c}}) + \pi_2 P_{2}^{n}(A_{n}),
\end{equation*}
where $\pi_1$ and $\pi_2$ are prior probabilities for $H_{1}$ and $H_{2}$ respectively and $A_{n}^{\mathrm{c}}$ represents the complement of $A_{n}$. It turns out that to minimise the probability of error 
$A_{n}$ should be chosen according to a maximum a posteriori decision rule; if $i$ is the maximiser of $\pi_{i} P_{i}(z^{n})$, then $H_{i}$ is chosen. The probability of error decays exponentially with $n$ where the exponent is the Chernoff information between $P_{1}$ and $P_{2}$.

We now relate this to privacy by means of an example. Suppose $X \in \{x_{1}, x_{2} \}$ is a binary private variable. An adversary accesses a series of random variables $Y_{1}, Y_{2}, \dots, Y_{n}$ that are i.i.d. given $X$. She conducts a Bayesian hypothesis test to determine $X$ where $H_{1}: P = Q_{Y|X=x_{1}}$ and $H_{2}: P = Q_{Y|X=x_{2}}$. 
Many privacy metrics compare the probability of an adversary making a correct guess before and after having made an observation. As an example, we will take min entropy leakage, a special case of g-leakage \cite{Alvim_g}\cite{Alvim_g2}. 
We can write
\begin{equation*}
    \mathcal{L}_{n} = \log \sum_{y^{n} \in \mathcal{Y}^{n}} Q_{Y^{n}} (y^{n}) \frac{\max_{x \in \{ x_{1},x_{2} \}} Q_{X|Y^{n}=y^{n}}(x)}{\max_{x \in \{ x_{1},x_{2} \} } Q_{X}(x)},
\end{equation*}
which can be rearranged to give\footnote{Alternatively, one could simply note that (\ref{eq: min entropy leakage operational def}) is the operational definition for min entropy leakage.}
\begin{equation} \label{eq: min entropy leakage operational def}
    \mathcal{L}_{n} = \log \frac{1 - P_{e}^{(n)}}{1-P_{e}},
\end{equation}
where $P_{e}$ is the adversary's probability of an erroneous guess without any observations. Table \ref{table: existing global metrics} shows that min entropy leakage is in the set of global privacy metrics defined in Section \ref{section: global leakage}. Thus by Theorem \ref{theorem: global rate} we know that it approaches its limit, $\mathcal{L}_{\infty}$, exponentially at a rate governed by the Chernoff information. By Corollary \ref{corollary: global data processing}, we know that this approach is from below and therefore that the asymptotic properties (\ref{eq: global asymptotic property 1}-\ref{eq: global asymptotic property 2}) are satisfied.
This can equally be argued using the result from Bayesian hypothesis testing. We know that $1-P_{e}^{(n)}$ tends to $1$ from below exponentially according to the Chernoff information. The final result follows directly from the Taylor series expansion of the logarithm. 

We can further intuit the case when $X$ is not binary. The adversary determines $X$ by pointwise comparison, testing all pairings $x, x'$. The error probability that decays the slowest and is therefore the only relevant term in the limit of large $n$ is that of the pair with the smallest Chernoff information. The exponent is $\min_{x \neq x'} \textcolor{black}{\mathscr{C}} \left( Q_{Y^{n}|X=x}|| Q_{Y^{n}|X=x'} \right)$, as we have found previously.

\section{Discussion and Conclusions}
We have discussed the essential asymptotic properties that an information leakage measure should satisfy. Namely, under the composition of a large number of observations for the adversary, leakage should approach its limit from below. We have outlined what this means for global and pointwise measures respectively, and used the arguments to motivate a set of axioms that a reasonable pointwise measure should follow. A set of corresponding global measures was defined which retains a great deal of  generality and flexibility, and encompasses many existing measures. 
Thus, any pointwise or global privacy metrics proposed in future may well satisfy \ref{item: A1}-\ref{item: A5} or fall within our global set respectively. The asymptotic properties of such metrics could straightforwardly be obtained from this paper. 
Future work may seek to generalise the global set further, in order to encompass all `reasonable' global information leakage metrics, analogously to the pointwise result.

For the given pointwise and global metrics, we provided composition theorems which state that privacy degrades exponentially according to the minimum Chernoff information between distinct pairs $Q_{Y|X=x}, Q_{Y|X=x'}$. Defined as a random variable, pointwise leakage tends almost surely to the random information contained in the private variable $X$. It is worth noting that a pointwise measure need not be a reasonable information leakage metric according to \ref{item: A1}-\ref{item: A5} to follow Theorem \ref{theorem: rate of L1 convergence}. Specifically, \ref{item: A1}-\ref{item: A4} need not be followed. This is because of the underlying fact that as $n \to \infty$ the posterior distribution of $X$ converges almost surely to one of the extreme points of the simplex, $E_i$. So, when the function is sufficiently smooth at the extreme points, Theorem 2 may hold. 
Whilst we say that a reasonable pointwise information leakage metric should follow all five axioms for the reasons detailed in Section \ref{section: axioms pointwise}, the subsequent analysis only requires \ref{item: A5}. Thus, the result encompasses but is not restricted to so called `reasonable pointwise information leakage metrics'. Future work could explore further applications of Theorem \ref{theorem: rate of L1 convergence}, as well as the consequences of \ref{item: A1}-\ref{item: A4} on the behaviour of a reasonable pointwise information leakage metric.

{\color{black} We showed that if a pointwise leakage function satisfies Condition~\ref{assumption: f double bound}, then the rate is governed by the minimum Chernoff information. This essentially follows from the way the function is behaving near the extreme points. 
In Condition~\ref{assumption: f double bound}, the function is behaving like $z \left( \log \frac{1}{z}\right)^{s}$ near the extreme point $E_i$ for some $s$, where $z = 1-p_i$. 
We can generalize this condition with arbitrary functions $S_u(z)$ and $S_l(z)$ in the upper and lower bounds, respectively. 
If a pointwise privacy function $f$ satisfies \eqref{eq: condition 1} 
with these functions (assuming them to be well-behaved), i.e., $S_l(1-p_i) \leq f(E_i, Q) - f(P,Q) \leq S_u(1-p_i)$, then it might be possible to show using our arguments that the rate at which the privacy degrades would be within $-C \cdot\lim_{z \to 0}\frac{\log S_l(z)}{\log z}$  and  $-C \cdot\lim_{z \to 0} \frac{\log S_u(z)}{\log z}$.
This relies on the fact that $z= 1-p_i$ behaves like $2^{-nC}$ around each $E_i$, as noted in Lemma~\ref{lemma: conditional dist to Ex}.}

Finally, we discussed the connection between privacy and Bayesian hypothesis testing. The takeaway point is that the mathematics underpinning composition theorems for privacy metrics can be understood via the setting of Bayesian hypothesis testing.
Underlying both is the method of types. This is because all of the measures are functions of $Q_{Y^n|X}$ where analysis of types naturally appears.

\section{Proofs of the Main Results}\label{section: proofs}

\subsection{Notation}
To follow the proofs, some further notation is required (most of which follows that of \cite{wu2020}). 
Base $2$ logarithms are assumed. 
It will be important to note that probabilities of random variables are themselves random variables. For example, $Q_{Y|X=x}(Y)$ is a random variable representing the probability of observing $Y$, the random output of $Q_{Y|X=x}$.

The argument,
\begin{equation*}
    x^{\star}(y^{n}) \defeq \arg \max_{x} Q_{Y^{n}|X=x}(y^{n}),
\end{equation*}
is the adversary's maximum likelihood estimate (MLE) for $x$. 
Where appropriate $x^{\star}$ and $X^{\star}$ are written as $x^{\star}(y^{n})$ and $x^{\star}(Y^{n})$ respectively to emphasise their dependence on the adversary's observations. 
The set $\mathcal{P}_n$ contains all possible empirical distributions corresponding to a length $n$ sequence over $\mathcal{Y}$; to each $y^n$, we can associate an empirical distribution $P_n \in \mathcal{P}_n$. 
Meanwhile, $\mathcal{P}$ is the set of all possible distributions over $\mathcal{Y}$, and any $P\in \mathcal{P}$ represents one of these possible distributions. Also let $T(P_{n})$ be the set of $y^{n} \in \mathcal{Y}^{n}$ whose empirical distribution is $P_{n}$. 
As it is frequently used, the distribution $Q_{Y|X=x}$ is shortened to $Q_{x}$. 
Next, let $x_{k}(P)$ represent the $x$ realisation that induces the $k$th smallest Kullback-Leibler divergence, $D(P||Q_{x})$, across all possible realisations $x \in \mathcal{X}$; i.e., $D(P||Q_{x_{1}(P)}) = \min_{x}D(P||Q_{x})$.\footnote{Whenever there is ambiguity in the ordering, we fix an order and work with it.} 
We also define $x$-domains as follows:
\begin{align*}
    D_{x} &\defeq \left\{ P \in \mathcal{P} | D(P||Q_{x}) < D(P||Q_{x'}) \; \forall x' \textcolor{black}{\neq x} \in \mathcal{X} \right\}, 
\end{align*}
For brevity, let
\begin{align*}
    C_{n} &\defeq \inf_{P_{n}\in \mathcal{P}_{n}} D(P_{n}||Q_{x_{2}(P_{n})}), \\
    C &\defeq \min_{x \neq x'} \textcolor{black}{\mathscr{C}} (Q_{x}||Q_{x'}). 
\end{align*}
We call the latter the minimum Chernoff information.

\subsection{Auxiliary Remarks and Lemmas}
We can make three remarks which will frequently be pertinent throughout the proofs.
\begin{remark} \label{remark: x1 = x*}
    If $P_{n}$ is the empirical distribution of $y^{n}$,
    \begin{equation*}
        x_{1}(P_{n}) = x^{\star}(y^{n}).
    \end{equation*}
          By noting  that $P_{n}(y)$ is the fraction of occurrences of $y$ in $y^n$, it follows that
    \begin{align*}
        x^{\star}(y^{n}) &= \arg \max_{x} Q_{Y^{n}|X=x}(y^{n}) \\
        &= \arg \max_{x} \prod_{y \in \mathcal{Y}} Q_{Y|X=x}(y)^{n P_{n}(y)} \\
        &= \arg\max_{x} \sum_{y \in \mathcal{Y}} P_{n}(y) \log Q_{Y|X=x}(y) \\
        &= \arg \min_{x} \sum_{y \in \mathcal{Y}} P_{n}(y) \log \frac{P_{n}(y)}{Q_{Y|X=x}(y)} \\
        &= \arg \min_{x} D(P_{n}||Q_{x})
        = x_{1}(P_{n}).
    \end{align*}
    We note that if there are multiple $x$ values with the same minimum $D(P_n||Q_x)$, these $x$ values also all have the same maximum $Q_{Y^{n}|X=x}(y^{n})$. In this case we set by convention that $x_1(P_n) = x^\star (y^n)$. 
\end{remark}
\begin{remark} \label{remark: P sets}
    By standard continuity arguments \textcolor{black}{for the KL divergence \cite[Sec. III-A]{KLcontinuous}},
    \begin{equation*}
        \lim_{n \to \infty} \inf_{P_n \in \mathcal{P}_n}D(P_n|| Q_{x_2(P_n)}) = \inf_{P \in \mathcal{P}}D(P||Q_{x_2(P)}),
    \end{equation*}
    as \textcolor{black}{the distributions are discrete and the set of \emph{all} empirical distributions is dense in a probability simplex. } 
\color{black}{It is obvious that  $\inf_{P_n \in \mathcal{P}_n}D(P_n|| Q_{x_2(P_n)}) \geq  \inf_{P \in \mathcal{P}}D(P||Q_{x_2(P)})$, which implies one direction of the equality. For the other direction, we may apply the continuity in the first argument of KL divergence \cite[Sec. III-A]{KLcontinuous} with $\hat{P} = \arg\inf_{P \in \mathcal{P}} D(P||Q_{x_2(P)})$ to a sequence of empirical distributions $P_k \to \hat{P}$ ($L^{1}$-norm) with $x_2(P_k) = x_2(\hat{P})$. The latter condition follows again from the continuity of the KL divergence around $\hat{P}$ and the fact that $x_2(P_k)$ is a discrete minimiser of it. This means that the expression on the LHS
of the equality is no larger than 
$\lim_{n \to \infty} D(P_k|| Q_{x_2(P_k)}) = \lim_{n \to \infty} D(P_k|| Q_{x_2(\hat{P})})= D(\hat{P}||Q_{x_2(\hat{P})})$, which is the RHS of the above equality.}
\end{remark}

\begin{remark} \label{remark: chernoff}
    It follows from Remark \ref{remark: P sets} and from the result of \cite[Lemma 4]{wu2020} which is
    \begin{equation*}
        \inf_{P \in \mathcal{P}} D(P||Q_{x_{2}(P)}) = \min_{x \neq x'}\textcolor{black}{\mathscr{C}}(Q_{x}||Q_{x'}),
    \end{equation*}
    that $\lim_{n \to \infty} C_{n} = C$, \textcolor{black}{and we also have that $C \leq C_n$}. 
\end{remark}

\begin{figure}[ht!]
  \centering
\includegraphics[scale=0.4]{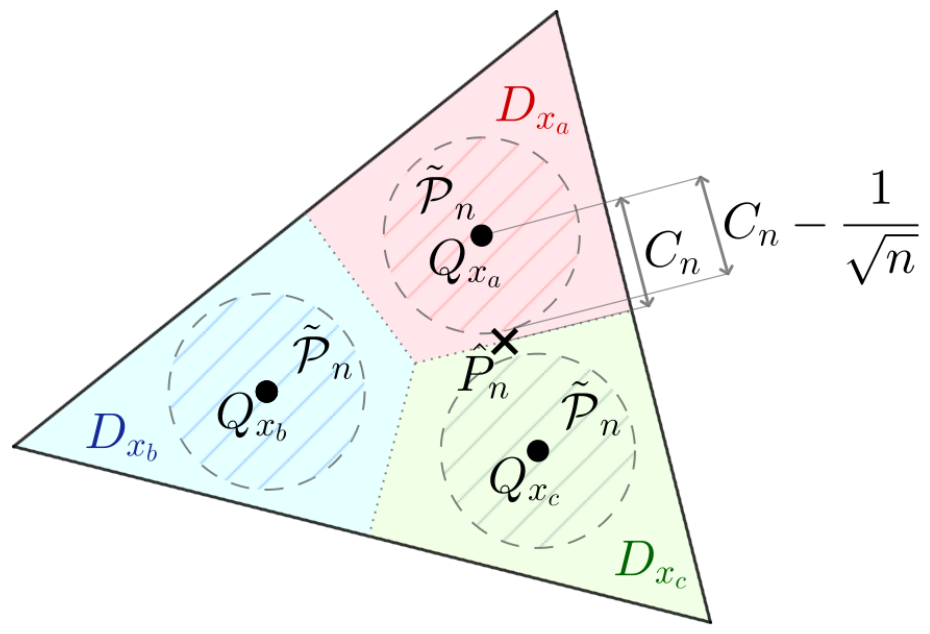}
  \caption{An example over the probability simplex $\mathcal{P}$ with $|\mathcal{Y}|=3$}
  \label{fig: geogebra simplex}
\end{figure}

Let us also define the set $\mathcal{\widetilde{P}}_{n} \subseteq \mathcal{P}_{n}$ as the set of $P_{n} \in \mathcal{P}_{n}$ satisfying\footnote{On the right hand side, any function from the class $\omega(1/n)$ could have been chosen.}
\begin{equation}
    C_{n} - D(P_{n}||Q_{x_{1}(P_{n})}) \geq \frac{1}{\sqrt{n}}. \label{eq: little omega condition}
\end{equation}
We denote the corresponding set of possible $y^{n}$ sequences as $\widetilde{\mathcal{Y}}_{n}$. The set can be visualised on a probability simplex, as shown in Figure \ref{fig: geogebra simplex},\footnote{Note that `differences' between points are measured as KL divergences, not geometric distances.} where $\hat{P}_{n}$ is the argument of $C_{n}$. 

We outline five lemmas in turn to prove Theorems \ref{theorem CDF convergence} and \ref{theorem: rate of L1 convergence}. 
Lemma \ref{lemma: not tilde prob} says that when $n$ is large, $y^{n}$ lies within $\widetilde{\mathcal{Y}}_{n}$ with high probability, which will allow us to restrict the majority of our analysis to the set. Lemma \ref{lemma: conditional dist to Ex} is the crux of the proof. 
It says that, with high probability, $Q_{X|Y^{n}=y^{n}}$ and $E_{x^{\star}}$ become close as $n$ grows large. 
Lemma \ref{lemma: part 1} makes use of Lemma \ref{lemma: conditional dist to Ex} to say that, for $y^{n} \in \widetilde{\mathcal{Y}}^{n}$, the pointwise leakage function approaches the information function of $x^{\star}(y^{n})$ from below with increasing $n$. 
In turn, Lemma \ref{lemma: part 2} says that the distribution of $x^{\star}(Y^{n})$ approaches the distribution of $X$. Finally, Lemma \ref{lemma: part 3 new} combines Lemmas \ref{lemma: part 1} and \ref{lemma: part 2} to bound the difference between the pointwise leakage CDF and the information CDF at any given point. 
Lemmas \ref{lemma: not tilde prob}-\ref{lemma: part 2} are used along with the Borel-Cantelli Lemma to prove Theorem \ref{theorem CDF convergence}. Lemma \ref{lemma: part 3 new} is used to prove Theorem \ref{theorem: rate of L1 convergence}. To prove Theorem \ref{theorem: global rate}, we introduce Lemma \ref{Lemma: global part 1}, which adapts Lemma \ref{lemma: part 1} for use in the global setting.

\begin{lemma} \label{lemma: not tilde prob}
    The probability of a sequence $y^{n}$ lying outside the set $\widetilde{\mathcal{Y}}_{n}$ decreases exponentially with $n$ at a rate governed by the minimum Chernoff information,
    \begin{equation*}
        \sum_{y^n \not\in \widetilde{\mathcal{Y}}_{n}}Q_{Y^{n}}(y^{n}) \leq \psi,
    \end{equation*}
    where
    \begin{equation*}
        \psi \defeq |\mathcal{X}| (n+1)^{|\mathcal{Y}|} 2^{-n\left(C - \frac{\textcolor{black}{1}}{\sqrt{n}}\right)},
    \end{equation*}
    \textcolor{black}{for all large enough $n$.}
    \begin{proof}
        \begin{align}
    \sum_{y^n \not\in \widetilde{\mathcal{Y}}_{n}}Q_{Y^{n}}(y^{n}) &= \sum_{P_{n} \not\in \widetilde{\mathcal{P}}_{n}} \sum_{x\in \mathcal{X}} Q_{Y^{n}|X=x}(T(P_{n})) Q_{X}(x) \nonumber \\
    &\leq \sum_{P_{n} \not\in \widetilde{\mathcal{P}}_{n}} \sum_{x\in \mathcal{X}} 2^{-n D(P_{n}||Q_{x})} \label{eq: prob not in set small l2} \\
    &\leq |\mathcal{X}| \sum_{P_{n} \not\in \widetilde{\mathcal{P}}_{n}} 2^{-nD(P_{n}||Q_{x_{1}(P_{n})})} \label{eq: prob not in set small l3} \\
   & \leq |\mathcal{X}| (n+1)^{|\mathcal{Y}|} 2^{-n\left(C_{n}-\frac{1}{\sqrt{n}}\right)} \label{eq: prob not in set small l4}\\
   & \leq |\mathcal{X}| (n+1)^{|\mathcal{Y}|} 2^{-n\left(C   - \frac{\textcolor{black}{1}}{\sqrt{n}}\right)}\label{eq: prob not in set small l5}
\end{align}
where (\ref{eq: prob not in set small l2}) follows from \cite[Th. 11.1.4]{Cover2006},
(\ref{eq: prob not in set small l3}) follows from the definition of $x_{1}(P)$, 
(\ref{eq: prob not in set small l4}) follows from the definition of $\widetilde{\mathcal{P}}_{n}$ (\ref{eq: little omega condition}), and
(\ref{eq: prob not in set small l5}) \textcolor{black}{uses $C \leq C_n$}.
    \end{proof}
\end{lemma}
\noindent As $C$ is constant, we can conclude from Lemma \ref{lemma: not tilde prob} that
\begin{equation}
    \lim_{n \to \infty} \sum_{y^n \not\in \widetilde{\mathcal{Y}}^{n}}Q_{Y^{n}}(y^{n}) = 0. \label{eq: small set prob}
\end{equation}

Lemma \ref{lemma: conditional dist to Ex} is the crux of the proof and says that, with high probability, the posterior distribution $Q_{X|Y^{n}=y^{n}}$ approaches $E_{x^{\star}}$ as $n$ grows large.
\begin{lemma} \label{lemma: conditional dist to Ex}
For all $P_{n} \in \widetilde{\mathcal{P}}_{n}$ and $n$ large enough,
        \begin{align}
            \frac{1}{2} \min_{x}Q_{X}(x)2^{-nK(P_{n})} \leq 1-Q_{X|Y^{n}=y^{n}}(x^{\star}) \leq \frac{1}{\min_{x}Q_{X}(x)} 2^{-nK(P_{n})}, \label{eq: lb lemma 1 x star prob bound}
        \end{align} 
        where
        \begin{equation*}
            K(P_{n}) \defeq D(P_{n}||Q_{x_{2}(P_{n})}) - D(P_{n}||Q_{x^{\star}}).
        \end{equation*}
    \begin{proof}
    First, consider the probability $Q_{X|Y^{n}=y^{n}}(x^{\star})$.
        \begin{align}
            Q_{X|Y^{n}=y^{n}}(x^{\star}) &= \frac{Q_{Y^{n}|X=x^\star}(y^{n})Q_{X}(x^{\star})}{Q_{Y^{n}}(y^{n})} \nonumber\\
            &= \frac{2^{-nD(P_{n}||Q_{x^{\star}})}Q_{X}(x^{\star})}{2^{-nD(P_{n}||Q_{x^{\star}})}Q_{X}(x^{\star}) + \sum_{x\neq x^{\star}}2^{-nD(P_{n}||Q_{x})}Q_{X}(x)} \nonumber \\
            &= \left( 1+ \frac{2^{nD(P_{n}||Q_{x^{\star}})}}{Q_{X}(x^{\star})}\sum_{x \neq x^{\star}}Q_{X}(x)2^{-nD(P_{n}||Q_{x})}\right) ^{-1}. \label{eq: lb elementwise probability l3}
        \end{align}
         To establish a lower bound, we can bound the summation.
    \begin{align*}
        \sum\limits_{x\neq x^{\star}}Q_{X}(x)2^{-n D(P_{n}||Q_{x})} &\leq  \sum\limits_{x'\neq x^{\star}}Q_{X}(x)\max\limits_{x\neq x^{\star}}2^{-n D(P_{n}||Q_{x})} \\
        & = \left(1-Q_{X}(x^{\star})\right) 2^{-n D(P_{n}||Q_{x_{2}(P_{n})})} \\
        & \leq 2^{-n D(P_{n}||Q_{x_{2}(P_{n})})}. 
    \end{align*}
       Substituting this into (\ref{eq: lb elementwise probability l3}) we find
        \begin{align*}
            Q_{X|Y^{n}=y^{n}}(x^{\star}) &\geq \left( 1+ \frac{1}{Q_{X}(x^{\star})}2^{-nK(P_{n})}\right) ^{-1} \\
            &\geq 1 - \frac{1}{Q_{X}(x^{\star})}2^{-nK(P_{n})} \\
            &\geq 1 - \frac{1}{\min_x Q_X(x)}2^{-nK(P_{n})}.
        \end{align*}
        Looking now for an upper bound, we continue from (\ref{eq: lb elementwise probability l3}).
        \begin{align*}
            Q_{X|Y^{n}=y^{n}}(x^{\star}) &\leq \left( 1+ \frac{2^{nD(P_{n}||Q_{x^{\star}})}}{Q_{X}(x^{\star})} Q_{X}(x_{2}(P_{n}))2^{-nD(P_{n}||Q_{x_{2}(P_{n})})}\right) ^{-1} \\
            &\leq 1 - \frac{Q_{X}(x_{2}(P_{n}))}{Q_{X}(x^{\star})} 2^{-nK(P_{n})} + 
            \frac{Q_{X}(x_{2}(P_{n}))^{2}}{Q_{X}(x^{\star})^{2}} 2^{-2nK(P_{n})} \\
            &\leq 1 - \min_x Q_X(x) 2^{-nK(P_{n})} + \frac{1}{\min_x Q_X (x)^2} 2^{-2nK(P_{n})}.
        \end{align*}
        As $P_{n} \in \widetilde{\mathcal{P}}_{n}$, $2^{-2nK(P_{n})}$ decays faster than $2^{-nK(P_{n})}$ and we can say that eventually,
        \begin{equation*}
             \frac{1}{2} \min_x Q_X(x) 2^{-nK(P_{n})} \geq \frac{1}{\min_x Q_X (x)^2} 2^{-2nK(P_{n})}.
        \end{equation*}
        Thus, for $n$ large enough,
        \begin{equation*}
            Q_{X|Y^{n}=y^{n}}(x^{\star}) \leq 1 - \frac{1}{2} \min_x Q_X(x) 2^{-nK(P_{n})}. 
        \end{equation*}
       Putting the results together we have 
        bounds on $E_{x^\star}(x^\star)-Q_{X|Y^n=y^n}(x^\star)=1- Q_{X|Y^n=y^n}(x^\star)$ as in
        (\ref{eq: lb lemma 1 x star prob bound}).
    \end{proof}
\end{lemma}

We next make use of Lemma \ref{lemma: conditional dist to Ex} to state Lemma \ref{lemma: part 1}.

\begin{lemma} \label{lemma: part 1}
    For any observation $y^{n}$ whose type $P_{n} \in \widetilde{\mathcal{P}}^{n}$, the pointwise leakage function is bounded by the information function as follows for large enough $n$. \textcolor{black}{For a pointwise leakage function $f$ that satisfies Condition \ref{assumption: f double bound},
    \begin{equation} \label{eq: lemma part 1 general}
        i_{X}(x^\star (y^n)) - \theta_1(n) 2^{-n K(P_n)} \leq l(y^n) \leq i_{X}(x^\star (y^n)),
    \end{equation}
    and for a pointwise leakage function that satisfies Condition \ref{assumption: f double bound} with $A>0$,
    \begin{equation} \label{eq: lemma part 1 derivative}
        i_{X}(x^{\star}(y^{n})) - \theta_{1}(n)2^{-nK(P_{n})} \leq l(y^n) \leq i_{X}(x^{\star}(y^{n})) -  \theta_{2}(n)2^{-nK(P_{n})},
    \end{equation}
    where
    \begin{align*}
        \theta_{1}(n) &\defeq B' n^{|b|} \left( 1 + O\left( \frac{1}{\sqrt{n}} \right) \right), \\
        \theta_{2}(n) &\defeq A' n^{-|a|} \left( 1 - O\left( \frac{1}{\sqrt{n}} \right) \right),
    \end{align*}
    for some strictly positive constants $B'$ and $A'$.
    \begin{proof}
    We begin with the case where $f$ satisfies Condition~\ref{assumption: f double bound}. It follows from Condition \ref{assumption: f double bound} that for the distribution $Q_{X}$, there exist neighbourhoods around the extreme points and constants $A\geq0$, $B>0$ and $a, b \in \mathbb{R}$ such that the pointwise leakage function $f$ satisfies \eqref{eq: condition 1}.
    When $n$ is large enough, Lemma \ref{lemma: conditional dist to Ex} and the fact that $K(P_n) \geq \frac{1}{\sqrt{n}}$ guarantees that $Q_{X|Y^n=y^n}$ will lie in the neighbourhood of $E_{x^\star}$ for every $P_n \in \widetilde{\mathcal{P}}_n $. Therefore, $l(y^n) \leq i_X(x^\star(y^n))$ follows immediately from the lower bound in \eqref{eq: condition 1} as $A\geq 0$. To prove the lower bound in \eqref{eq: lemma part 1 general}, we use the definition of Condition \ref{assumption: f double bound}:  
    \begin{align}
        i_X(x^\star&(y^n)) - l(y^n)\nonumber\\ &\leq B (1- Q_{X|Y^n=y^n}(x^\star) ) \left( \log \frac{1}{1- Q_{X|Y^n=y^n}(x^\star)}\right)^{b} \nonumber \\
        &\leq \frac{B}{\min_x Q_X(x)} 2^{-n K(P_n)} \left( \log \frac{2}{\min_xQ_X(x) 2^{-nK(P_n)}}\right)^{b} \label{eq: lemma3 apply lemma 2} \\
        & \leq \frac{B}{\min_x Q_X(x)} 2^{-n K(P_n)} \left( \log \frac{2}{\min_xQ_X(x) 2^{-nK(P_n)}}\right)^{|b|} \label{eq: mod b} \\
        &= \frac{B}{\min_x Q_X(x)} 2^{-nK(P_n)} n^{|b|} K(P_n)^{|b|} \left( 1 + \frac{1}{nK(P_n)} \log\frac{2}{\min_x Q_X(x)} \right)^{|b|}\nonumber \\
        & \leq \frac{B}{\min_x Q_X(x)} 2^{-nK(P_n)} n^{|b|} K(P_n)^{|b|} \left(1 +  \frac{2|b|}{nK(P_n)} \log\frac{2}{\min_x Q_X(x)}\right) \label{eq: lemma 3 using n large} \\
        & \leq \frac{B}{\min_x Q_X(x)} 2^{-nK(P_n)} n^{|b|} K(P_n)^{|b|} \left(1 + \frac{2|b|}{\sqrt{n}} \log\frac{2}{\min_x Q_X(x)} \right) \label{eq: lemma 3 use K} \\
        &\leq \frac{B}{\min_x Q_X(x)} 2^{-nK(P_n)} n^{|b|} (2C)^{|b|} \left(1 + \frac{2|b|}{\sqrt{n}} \log\frac{2}{\min_x Q_X(x)} \right) \label{eq: use Cn C remark} \\
        &= B' n^{|b|} \left( 1+ O\left(\frac{1}{\sqrt{n}} \right) \right) 2^{-nK(P_n)},  \label{eq: B bound}
    \end{align}
    where the constant $B' := \frac{B (2C)^{|b|}}{\min_x Q_X(x)}>0$ and $C$ is the minimum Chernoff information, (\ref{eq: lemma3 apply lemma 2}) applies Lemma \ref{lemma: conditional dist to Ex}, (\ref{eq: mod b}) assumes $n$ is large enough that $\log \frac{2}{\min_xQ_X(x) 2^{-nK(P_n)}} \geq 1$ as $K(P_n) \geq \frac{1}{\sqrt{n}}$ for $P_n \in \widetilde{\mathcal{P}}_n$, (\ref{eq: lemma 3 using n large}) assumes $n$ is large enough and makes use of the inequality $(1+z)^{|b|} \leq 1+2|b|z$, which holds when $0\leq z \leq 2^{\frac{1}{|b|-1}}-1$ if $|b| >1$ and $0 \leq z \leq 1$ if $|b|\leq 1$, and (\ref{eq: lemma 3 use K}) uses $K(P_n) \geq \frac{1}{\sqrt{n}}$ for $P_n \in \widetilde{\mathcal{P}}_n$.
    To bound $K(P_n)^{|b|}$ in (\ref{eq: use Cn C remark}), we have used $ K(P_n) \leq C_n$ and Remark \ref{remark: chernoff} to say $C_n \leq 2C$ for large enough $n$ as $\lim_{n \to \infty}C_n=C$. We have reached the lower bound in (\ref{eq: lemma part 1 general}). \\
    Let us now assume that the pointwise leakage function $f$ satisfies Condition \ref{assumption: f double bound} with $A>0$. Again, for every $P_n \in \widetilde{\mathcal{P}}_n$ and $n$ large enough, Lemma \ref{lemma: conditional dist to Ex} and the fact that $K(P_n) \geq \frac{1}{\sqrt{n}}$ guarantees that $Q_{X|Y^n=y^n}$ will lie in the corresponding neighbourhood of $E_{x^\star}$, over which \eqref{eq: condition 1} is satisfied. As Condition \ref{assumption: f double bound} with $A\geq 0$ is more general than with the restriction $A>0$, the above steps for \eqref{eq: B bound} are still valid, yielding the lower bound in (\ref{eq: lemma part 1 derivative}). The proof of the upper bound is similar:
    \begin{align}
        i_X(x^\star&(y^n)) - l(y^n)\\ \nonumber &\geq A (1- Q_{X|Y^n=y^n}(x^\star) ) \left( \log \frac{1}{1- Q_{X|Y^n=y^n}(x^\star)}\right)^{a} \nonumber \\
        &\geq \frac{A}{2} \min_x Q_X(x)  2^{-nK(P_n)} \left( \log \frac{\min_x Q_X(x)}{2^{-nK(P_n)}} \right)^{a} \label{eq: lemma3 apply lemma 2, 2} \\
        &\geq \frac{A}{2} \min_x Q_X(x)  2^{-nK(P_n)} \left( \log \frac{\min_x Q_X(x)}{2^{-nK(P_n)}} \right)^{-|a|} \label{eq: mod a} \\
        &= \frac{A}{2} \min_x Q_X(x) 2^{-nK(P_n)} n^{-|a|} K(P_n)^{-|a|} \left( 1 - \frac{1}{nK(P_n)} \log\frac{1}{\min_x Q_X(x)} \right)^{-|a|} \nonumber \\
        & \geq \frac{A}{2} \min_x Q_X(x) 2^{-nK(P_n)} n^{-|a|} K(P_n)^{-|a|} \left(1 - \frac{2|a|}{nK(P_n)} \log\frac{1}{\min_x Q_X(x)} \right) \label{eq: lemma 3 using n large, 2} \\
        & \geq \frac{A}{2} \min_x Q_X(x) 2^{-nK(P_n)} n^{-|a|} (2C)^{-|a|} \left(1 - \frac{2|a|}{\sqrt{n}} \log\frac{1}{\min_x Q_X(x)} \right) \label{eq: lemma 3 use K, 2} \\
        &= A' n^{-|a|}  \left( 1- O \left(\frac{1}{\sqrt{n}} \right) \right)2^{-nK(P_n)}, \nonumber
    \end{align}
    where $A' = \frac{A \min_x Q_X(x) (2C)^{-|a|}}{2} > 0$, (\ref{eq: lemma3 apply lemma 2, 2}) applies Lemma \ref{lemma: conditional dist to Ex}, (\ref{eq: mod a}) assumes $n$ is large enough that $\log \frac{\min_x Q_X(x)}{2^{-nK(P_n)}}  \geq 1$ as $K(P_n) \geq \frac{1}{\sqrt{n}}$ for $P_n \in \widetilde{\mathcal{P}}_n$, (\ref{eq: lemma 3 using n large, 2}) assumes $n$ is large enough and makes use of the inequality $(1-z)^{-|a|} \geq 1-2|a|z$, which holds when $0\leq z \leq 1$ as $-|a|\leq 0$, and (\ref{eq: lemma 3 use K, 2}) uses $K(P_n)\leq 2C$ for $P_n \in \widetilde{\mathcal{P}}_n$ when $n$ is sufficiently large. This completes the proof of (\ref{eq: lemma part 1 derivative}).
    \end{proof}}
\end{lemma}

\textcolor{black}{When $n$ is large enough, the exponential term decays faster than the polynomial terms of $\theta_1(n)$ and $\theta_2(n)$ in the above lemma. Thus}
we have shown that, with high probability, the pointwise leakage function tends to the information function evaluated at $X=x^{\star}$. For the pointwise leakage CDF to tend to the information CDF, it must therefore be the case that the distribution of $X^{\star} = x^{\star}(Y^{n})$ tends to the distribution of $X$. We state this in Lemma \ref{lemma: part 2}.

\begin{lemma} \label{lemma: part 2}
    The distribution of $X^{\star}$ is related to the distribution of $X$ as follows
    \begin{equation} \label{eq: lemma 4 statement 1}
        \big| Q_{Y^{n}}(S_{x}) - Q_{X}(x) \big| \leq \phi,
    \end{equation}
    and
    \begin{equation} \label{eq: lemma 4 statement 2}
        \Pr\left\{ x^{\star}(Y^{n}) \neq X \right\} \leq \phi,
    \end{equation}
    for any $x \in \mathcal{X}$, where
    \begin{align*}
        S_{x} &\defeq \left\{y^{n} \in \mathcal{Y}^{n} : x^{\star}(y^{n})=x \right\} \\
        \phi &\defeq (n+1)^{|\mathcal{Y}|}2^{-nC_{n}}.
    \end{align*}
    \begin{proof}
    Starting with the lower bound of (\ref{eq: lemma 4 statement 1}),
    \begin{align}
        Q_{Y^{n}}(S_{x}) &= \sum_{x' \in \mathcal{X}}  Q_{Y^{n}|X=x'}(S_{x})Q_{X}(x') \nonumber \\ 
        &\geq \sum_{P_{n} \in \mathcal{P}_{n} \cap D_{x}} \sum_{x' \in \mathcal{X}} Q_{Y^{n}|X=x'}(T(P_{n}))Q_{X}(x') \label{eq: part 2 LB l2} \\
        & \geq \sum_{P_{n} \in \mathcal{P}_{n} \cap D_{x}} Q_{Y^{n}|X=x}(T(P_{n}))Q_{X}(x) \label{eq: part 2 LB l3} \\
        &= Q_{X}(x) \left[ 1 - \sum_{P_{n} \in \mathcal{P}_{n} \setminus D_{x}} Q_{Y^{n}|X=x}(T(P_{n})) \right] \nonumber \\
        & \geq Q_{X}(x) \left[ 1 - \sum_{P_{n} \in \mathcal{P}_{n} \setminus D_{x}} 2^{-n D(P_{n}||Q_{x})} \right], \label{eq: part 2 LB l6}
    \end{align}
    where in (\ref{eq: part 2 LB l2}) we recognise that $y^{n} \in S_{x} \iff x^{\star}(y^{n})=x \Leftarrow P_{n} \in D_{x}$ by Remark \ref{remark: x1 = x*} and the definition of  $D_{x}$. The last line follows from \cite[Theorem 11.1.4]{Cover2006}. Noting also that, if $P_{n}$ is not in the domain $D_x$,
    \begin{equation*}
        D(P_{n}||Q_{x}) \geq D(P_{n}||Q_{x_{2}(P_{n})}),
    \end{equation*}
    and we can bound the summation
    \begin{align*}
        \sum_{P_{n} \in \mathcal{P}_{n} \setminus D_{x}} 2^{-n D(P_{n}||Q_{x})} &\leq \sum_{P_{n} \in \mathcal{P}_{n} \setminus D_{x}} 2^{-n D(P_{n}||Q_{x_{2}(P_{n})})} \\ &\leq |\mathcal{P}_{n}| \max_{P_{n} \in \mathcal{P}_{n} \setminus D_{x}} 2^{-n D(P_{n}||Q_{x_{2}(P_{n})})} \\ 
        &\leq (n+1)^{|\mathcal{Y}|} 2^{-C_{n}}.
    \end{align*}
    Following from (\ref{eq: part 2 LB l6}):
    \begin{align}
        Q_{Y^{n}}(S_{x}) \geq Q_{X}(x) - (n+1)^{|\mathcal{Y}|}2^{-nC_{n}}. \label{eq: lemma part 2 LB final}
    \end{align}
We next examine the upper bound.
    \begin{align}
        Q_{Y^{n}}(S_{x}) &= \sum_{P_{n} \in 
        \mathcal{P}_{n}: y^n \in S_x} \sum_{x' \in \mathcal{X}} Q_{Y^{n}|X=x'}(T(P_{n}))Q_{X}(x') \nonumber \\
        &= Q_{X}(x) \sum_{P_{n} \in 
        \mathcal{P}_{n}: y^n \in S_x} Q_{Y^{n}|X=x}(T(P_{n})) + \sum_{x' \neq x}Q_{X}(x') \sum_{P_{n} \in 
        \mathcal{P}_{n}: y^n \in S_x} Q_{Y^{n}|X=x'}(T(P_{n})) \nonumber \\
        &\leq Q_X (x) + \sum_{x' \neq x}Q_{X}(x') \sum_{P_{n} \in 
        \mathcal{P}_{n}: y^n \in S_x} 2^{-n D(P_n || Q_{x'})} \nonumber \\
        &\leq Q_{X}(x) + (1-Q_{X}(x)) \sum_{P_{n} \in 
        \mathcal{P}_{n}: y^n \in S_x} 2^{-n D(P_{n}||Q_{x_{2}(P_{n})})} \label{eq: lemma 4 ub l4} \\
        &\leq  Q_{X}(x) + |\mathcal{P}_{n}| \max_{P_{n} \in \mathcal{P}_n} 2^{-n D(P_{n}||Q_{x_{2}(P_{n})})} \nonumber \\
        &\leq Q_{X}(x) + (n+1)^{|\mathcal{Y}|} 2^{-nC_{n}}, \nonumber
    \end{align}
    where (\ref{eq: lemma 4 ub l4}) follows from the fact that $x'$ cannot be $x_{1}(P_n)$ because $P_n$ is defined such that $y^n \in S_x$, and $x' \neq x$.
    Combining upper and lower bounds, we arrive at (\ref{eq: lemma 4 statement 1}). Proof of (\ref{eq: lemma 4 statement 2}) is similar. We can say
    \begin{align*}
        \Pr\left\{ x^{\star}(Y^{n}) = X \big| X=x \right\} &= \sum_{P_{n} \in 
        \mathcal{P}_{n}: y^n \in S_x} Q_{Y^{n}|X=x}(T(P_{n})) \\
        &\geq \sum_{P_n \in \mathcal{P}_{n} \cap D_x} Q_{Y^{n}|X=x}(T(P_{n})).
    \end{align*}
    Following steps (\ref{eq: part 2 LB l3}-\ref{eq: lemma part 2 LB final}), we arrive at
    \begin{align*}
        \Pr\left\{ x^{\star}(Y^{n}) = X \big| X=x \right\} \geq 1- (n+1)^{|\mathcal{Y}|}2^{-nC_{n}}.
    \end{align*}
    Finally,
    \begin{align*}
        \Pr\left\{ x^{\star}(Y^{n}) = X \right\} &= \sum_{x \in \mathcal{X}} \Pr\left\{ x^{\star}(Y^{n}) = X \big| X=x \right\}Q_{X}(x) \\
        &\geq 1- (n+1)^{|\mathcal{Y}|}2^{-nC_{n}}.
    \end{align*}
    which gives (\ref{eq: lemma 4 statement 2}).
    \end{proof}
\end{lemma}

Lemma \ref{lemma: part 3 new} combines Lemmas \ref{lemma: part 1} and \ref{lemma: part 2} to bound the difference between the pointwise leakage CDF and the information CDF at any given point.

\begin{lemma} \label{lemma: part 3 new}
\textcolor{black}{If leakage is defined such that $f$ satisfies Condition \ref{assumption: f double bound}, then}
    for any $l \in \mathbb{R}$, the pointwise leakage CDF is bounded by the information CDF as follows
    \begin{equation} \label{eq: lemma part 3}
        \big| F_{L_{n}}(l) - F_{I_{X}}(l) \big| \leq |\mathcal{X}|\phi + \psi + A_{n}(l),
    \end{equation}
    if $n$ is large enough, where
    \begin{align*}
        A_{n}(l) \defeq \sum_{y^{n} \in \widetilde{\mathcal{Y}}^{n}} Q_{Y^{n}}(y^{n}) \mathds{1}\left[ l < i_{X}(x^{\star}(y^{n})) \leq l + \textcolor{black}{\theta_1(n)} 2^{-nK(P_{n})} \right].
    \end{align*}
\begin{proof}
    Starting with the lower bound of (\ref{eq: lemma part 3}) and using $\mathds{1}[\cdot]$ as the indicator function,
    \begin{align}
        \Pr\{ L_{n} \leq l \} 
        &\geq \sum_{y^{n}\in \widetilde{\mathcal{Y}}^{n}} Q_{Y^{n}}(y^{n}) \mathds{1}\left[ l(y^n) \leq l \right] \nonumber \\
        &\geq \sum_{y^{n}\in \widetilde{\mathcal{Y}}^{n}} Q_{Y^{n}}(y^{n}) \mathds{1}\left[ i_{X}(x^\star (y^n)) \leq l \right] \label{eq: part 3 LB l3} \\
        &= \sum_{x} \sum_{\substack{y^n \in \widetilde{\mathcal{Y}}^{n}: \\ x^{\star} (y^n) = x}} Q_{Y^{n}}(y^{n}) \mathds{1}\left[ i_{X}(x) \leq l \right] \nonumber \\
        &= \sum_{x} Q_{Y^{n}}(S_{x}) \mathds{1}\left[ i_{X}(x) \leq l \right] -  \sum_{x} \sum_{\substack{y^n \notin \widetilde{\mathcal{Y}}^{n}: \\ x^{\star} (y^n) = x}}Q_{Y^{n}}(y^{n}) \mathds{1}\left[ i_{X}(x) \leq l \right] , \nonumber
    \end{align}
    where (\ref{eq: part 3 LB l3}) makes use of Lemma \ref{lemma: part 1}. 
    Continuing directly from above,
    \begin{align}
        \Pr\{ L_{n} \leq l \} 
        &\geq \sum_{x} \left( Q_{X}(x) - \phi \right) \mathds{1}\left[ i_{X}(x) \leq l \right] -  \sum_{y^{n}\notin \widetilde{\mathcal{Y}}^{n}}Q_{Y^{n}}(y^{n}) \label{eq: part 3 LB l6} \\
        &\geq \sum_{x} Q_{X}(x) \mathds{1}\left[ i_{X}(x) \leq l \right] - \phi |\mathcal{X}| - \psi \label{eq: part 3 LB l7}\\
        &= \Pr\{I_{X} \leq l \} - \phi |\mathcal{X}| - \psi, \nonumber
    \end{align}
where (\ref{eq: part 3 LB l6}) and (\ref{eq: part 3 LB l7}) make use of Lemmas \ref{lemma: part 2} and \ref{lemma: not tilde prob} respectively. Now bounding from above,
\begin{align}
     &\Pr\{ L_{n} \leq l \}  = \sum_{y^{n} \in \widetilde{\mathcal{Y}}^{n}} Q_{Y^{n}}(y^{n}) \mathds{1}\left[ l(y^n) \leq l \right] + \sum_{y^{n} \notin \widetilde{\mathcal{Y}}^{n}} Q_{Y^{n}}(y^{n}) \mathds{1} \left[ l(y^n) \leq l \right] \nonumber \\
     &\leq \sum_{y^{n} \in \widetilde{\mathcal{Y}}^{n}} Q_{Y^{n}}(y^{n}) \mathds{1}\left[ i_{X}(x^{\star}(y^{n})) \leq l + \textcolor{black}{\theta_1(n)} 2^{-nK(P_{n})} \right] +  \sum_{y^{n} \notin \widetilde{\mathcal{Y}}^{n}} Q_{Y^{n}}(y^{n}) \label{eq: part 3 UB l1} \\
     &\leq \sum_{y^{n} \in \widetilde{\mathcal{Y}}^{n}} \!\! Q_{Y^{n}}(y^{n}) \mathds{1}\left[ i_{X}(x^{\star}(y^{n})) \leq l \right] +\!\!\! \sum_{y^{n} \in \widetilde{\mathcal{Y}}^{n}} \!\!\! Q_{Y^{n}}(y^{n}) \mathds{1}\left[l < i_{X}(x^{\star}(y^{n})) \leq l + \textcolor{black}{\theta_1(n)} 2^{-nK(P_{n})} \right] + \psi \nonumber \\
     &= \sum_{x \in \mathcal{X}} \sum_{\substack{y^n \in \widetilde{\mathcal{Y}}^{n}: \\ x^{\star} (y^n) = x}} Q_{Y^n}(y^n) \mathds{1}\left[ i_X (x) \leq l \right]+ A_n (l) + \psi \nonumber \\
    &\leq \sum_{x \in \mathcal{X}} Q_{Y^{n}}(S_{x}) \mathds{1}\left[ i_{X}(x)\leq l \right] + A_n (l) + \psi
    \\ &\leq \sum_{x \in \mathcal{X}} \left( Q_{X}(x) + \phi \right) \mathds{1}\left[ i_{X}(x) \leq l \right] \label{eq: apply lemma 2} + A_{n}(l) + \psi 
    \\ &\leq \Pr \{I_{X} \leq l \} + \phi |\mathcal{X}| + A_{n}(l) + \psi, \nonumber
\end{align}
where (\ref{eq: part 3 UB l1}) applies (\ref{eq: lemma part 1 general}) of Lemma \ref{lemma: part 1}. 
Combining this with the lower bound gives Lemma \ref{lemma: part 3 new}.
\end{proof}
\end{lemma}

\subsection{Pointwise Leakage Proofs}
\subsubsection{Proof of Theorem \ref{theorem CDF convergence}} \label{section: proof of th 1}

To prove Theorem \ref{theorem CDF convergence}, we will show that that for a fixed $\epsilon$ the probability of the event $B_{n, \epsilon}$ that $i_X(X) -l(y^n)<0$ or $i_X(X)-l(y^n)> \epsilon$ is exponentially small for all $n$ large enough.
We will then make use of the Borel-Cantelli Lemma to say that with probability one, $l(y^n)$ is eventually bounded above by $i_X(X)$, and $l(y^n)$ approaches $i_X(X)$. We can rewrite the probability of the above event as
\begin{align}
    &\hspace{-0.5em}\Pr\{B_{n, \epsilon}\}  \leq \Pr \{i_X(X)-l(y^n)<0, x^\star (Y^n)=X \} +\Pr\{i_X(X)-l(y^n) > \epsilon, x^\star (Y^n)=X \} \nonumber\\& \qquad\qquad\qquad\qquad\qquad\qquad\qquad\qquad\qquad\qquad\qquad\qquad +\Pr\{x^\star(Y^n)\neq X\}  \nonumber \\
    &\hspace{-0.5em}\leq \Pr\{i_X(x^\star(Y^n))-l(y^n)<0 \} + \Pr\{i_X(x^\star(Y^n))-l(y^n)> \epsilon\} + \Pr\{x^\star (Y^n)\neq X \}, \label{eq: borel cantelli proof 2 probabilities}
\end{align}
where the first inequality follows from the fact that all the three events are disjoint.
We bound the first two terms of (\ref{eq: borel cantelli proof 2 probabilities}). Recall that Lemma \ref{lemma: part 1} applies to all $y^{n} \in \widetilde{\mathcal{Y}}^{n}$, i.e., all $y^{n}$ with type $P_{n} \in \widetilde{\mathcal{P}}_{n}$. 
From (\ref{eq: lemma part 1 general}) of Lemma \ref{lemma: part 1}, we can say that for all $y^{n} \in \widetilde{\mathcal{Y}}^{n}$,
\begin{equation*}
    0 \leq i_{X}(x^{\star}(y^{n})) - l(y^n) \leq \textcolor{black}{\theta_1(n)} 2^{-nK(P_{n})},
\end{equation*}
if $n$ is large enough. Note that for all $y^{n} \in \widetilde{\mathcal{Y}}^{n}$, $K(P_{n}) \geq 1/\sqrt{n}$. \textcolor{black}{As a result, the RHS $\theta_1(n) 2^{-nK(P_{n})}$ decays with increasing $n$}. For a fixed $\epsilon>0$, we can always choose $n$ such that the RHS does not exceed $\epsilon$. Therefore, for any $\epsilon > 0$
\begin{align}
    \Pr\Big\{ 0 \leq i_{X}(x^{\star}(Y^{n})) - l(y^n) \leq \epsilon \Big\} &\geq \Pr\left\{Y^{n} \in \widetilde{\mathcal{Y}}^{n} \right\}, \label{eq: thrm 1 proof big prob}
\end{align}
if $n$ is large enough. 
Employing Lemma \ref{lemma: not tilde prob} and using (\ref{eq: thrm 1 proof big prob}), we can bound the first two terms of (\ref{eq: borel cantelli proof 2 probabilities}).
\begin{align*}
    \Pr \{i_X(x^\star(Y^n))-l(y^n)<0 \} +\Pr  &\{ i_X(x^\star(Y^n))-l(y^n) > \epsilon \} \nonumber \\ &= 1 - \Pr \{ 0 \leq i_X(x^\star(Y^n)) - l(y^n) \leq \epsilon \} \\ & \leq \Pr \{Y^n \notin  \widetilde{\mathcal{Y}}^n \} \\
    &\leq |\mathcal{X}|(n+1)^{|\mathcal{Y}|} 2^{-n\left(C  -\frac{\textcolor{black}{1}}{\sqrt{n}} \right)} . 
\end{align*}
Lemma \ref{lemma: part 2} bounds the third term of (\ref{eq: borel cantelli proof 2 probabilities}). Putting these together we can make a statement. For any given constant $\epsilon > 0$, there exists $n'$ such that for all $n \geq n'$ the following holds:
\begin{align}\label{eq: event}
    \Pr\{B_{n, \epsilon} \} \leq (n+1)^{|\mathcal{Y}|}\left( |\mathcal{X}| +1 \right) 2^{-n \left( C - \frac{\textcolor{black}{1}}{\sqrt{n}}\right)}.
\end{align}
Consider the event described by the LHS of \eqref{eq: event}. Over all $n$ from $1$ to $n'$, the maximum number of occurrences of the event is $n'$, which is finite. Summing the RHS of \eqref{eq: event} over all $n$ from $1$ to $\infty$ yields a finite quantity.\footnote{D'Alembert's ratio test gives a ratio of $2^{-C}$.} Therefore, over all $n$ from $n'$ to $\infty$, the expected number of occurrences of the event is finite. 
Therefore, by the Borel-Cantelli lemma, for any $\epsilon>0$, $B_{n, \epsilon}$'s occur infinitely often with probability zero. So, for any $\epsilon$,  $0 \leq i_X(X)-l(y^n)<\epsilon $ happens eventually with probability one. We can thus say that $l(y^n)$ converges almost surely to $i_{X}(X)$ from below. This concludes the proof of Theorem \ref{theorem CDF convergence}.

\subsubsection{Proof of Theorem \ref{theorem: rate of L1 convergence}} \label{section: proof of th2}
We can use Lemma \ref{lemma: part 3 new} to prove the upper bound of Theorem \ref{theorem: rate of L1 convergence}. We know
\begin{equation*}
    || F_{L_{n}} - F_{I_{X}} ||_{1} \defeq \int_{l_{\min}}^{l_{\max}} \left| F_{L_{n}}(l) - F_{I_{X}}(l) \right| \mathrm{d}l,
\end{equation*}
where $l_{\max} = \max_{P} f(P,Q_X)$ and $l_{\min} = \min_{P} f(P,Q_X)$.
Note that the support is a finite interval. 
We can now bound the integral.
\begin{align}
    \int_{l_{\min}}^{l_{\max}} \left| F_{L_{n}}(l) - F_{I_{X}}(l) \right| \mathrm{d}l &\leq \int_{l_{\min}}^{l_{\max}} \left( |\mathcal{X}|\phi + \psi + A_{n} \right) \mathrm{d}l \nonumber \\
    & = \left(|\mathcal{X}|\phi + \psi \right) \left( l_{\max} - l_{\min} \right) + \textcolor{black}{\theta_1(n)} \sum_{y^{n} \in \widetilde{\mathcal{Y}}^{n}} Q_{Y^{n}}(y^{n}) 2^{-nK(P_{n})} . \label{eq: th2 proof UB l3}
\end{align}
Consider the second term:
\begin{align}
    \textcolor{black}{\theta_1(n)} \sum_{y^{n} \in \widetilde{\mathcal{Y}}^{n}} Q_{Y^{n}}(y^{n})  2^{-nK(P_{n})}
    &= \textcolor{black}{\theta_1(n)} \sum_{P_{n}\in \widetilde{\mathcal{P}}_{n}} \sum_{x \in \mathcal{X}} Q_{Y^{n}|X=x}(T(P_{n})) Q_{X}(x) 2^{-nK(P_{n})} \nonumber \\
    &\leq \textcolor{black}{\theta_1(n)} \sum_{P_{n}\in \widetilde{\mathcal{P}}_{n}} \sum_{x \in \mathcal{X}} 2^{-nD(P_{n}||Q_{x})}2^{-nD(P_{n}||Q_{x_{2}(P_{n})})}2^{nD(P_{n}||Q_{x^{\star}})} \nonumber \\
    & \leq \textcolor{black}{\theta_1(n)} \sum_{P_{n}\in \widetilde{\mathcal{P}}_{n}} \sum_{x \in \mathcal{X}}2^{-nD(P_{n}||Q_{x_{2}(P_{n})})} \label{eq: th 2 proof ub l6} \\ 
    &\leq \textcolor{black}{\theta_1(n)}  (n+1)^{|\mathcal{Y}|}|\mathcal{X}| 2^{-n C_{n}}, \label{eq: th2 UB proof l6}
\end{align}
where in (\ref{eq: th 2 proof ub l6}) we recognise that $D(P_n || Q_{x^{\star}}) \leq D(P_n || Q_x )$ for any $x \in \mathcal{X}$.
Combining \eqref{eq: th2 UB proof l6} with (\ref{eq: th2 proof UB l3}) \textcolor{black}{and Remark~\ref{remark: chernoff}, where $\phi = (n+1)^{|\mathcal{Y}|}2^{-nC_{n}}$ and $\psi = |\mathcal{X}| (n+1)^{|\mathcal{Y}|} 2^{-n\left(C - \frac{1}{\sqrt{n}}\right)}$}, we arrive at:
\begin{align*} 
   ||F_{L_{n}} - F_{I_{X}} ||_{1}  \leq |\mathcal{X}| (n+1)^{|\mathcal{Y}|} \left( 2 \left( l_{\max} - l_{\min} \right) + \textcolor{black}{\theta_1(n)} \right) 2^{-n \left(C - \frac{\textcolor{black}{1}}{\sqrt{n}}\right)}.
   \end{align*}
\textcolor{black}{As $\lim_{n \to \infty} \frac{1}{n}\log \left(|\mathcal{X}| (n+1)^{|\mathcal{Y}|} \left( 2 \left( l_{\max} - l_{\min} \right) + \theta_1(n) \right)\right)=0$, we have}
\begin{align*}
   &\textcolor{black}{\lim_{n \to \infty} \frac{1}{n} \log ||F_{L_{n}} - F_{I_{X}} ||_{1} \leq -C}.
\end{align*}
We have reached the upper bound,
and have the first part of Theorem \ref{theorem: rate of L1 convergence}, given by (\ref{eq: th 2 general}).

Next, we prove the lower bound for pointwise information leakage functions that satisfy \textcolor{black}{Condition \ref{assumption: f double bound} with $A>0$}. Note that for general pointwise information leakage functions, the lower bound of the $L^1$ norm is trivially zero.

\begin{figure}[ht!]
  \centering
\includegraphics[scale=0.6]{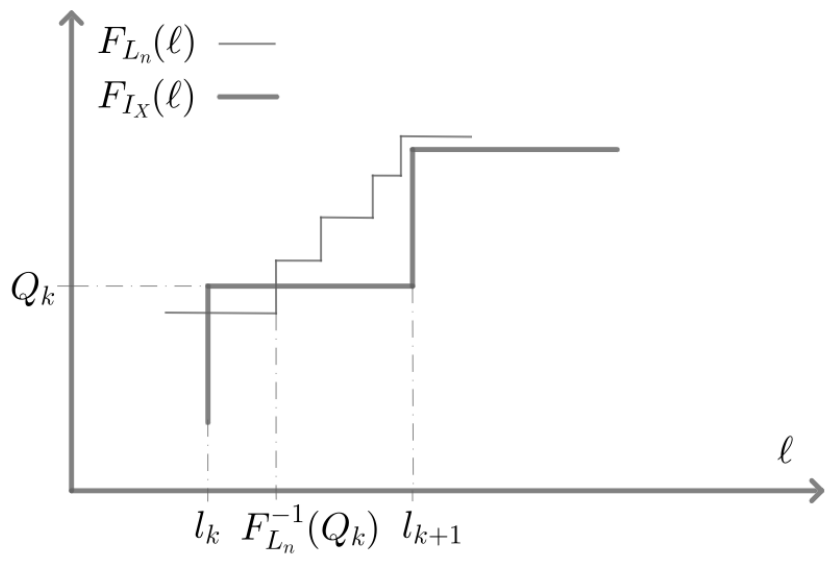}
  \caption{Example section of CDF plots}
  \label{fig: geogebra stairs}
\end{figure}

\textcolor{black}{First, we define information values as the set of all possible values of $i_X(x)$, of which there are as many as there are distinct values for $P_X(x)$.  Let us consider a partition $l_{\min}=l_1 < l_2< \cdots < l_s= l_{\max}$ of the interval $[l_{\min}, l_{\max}]$ using the information values. Here all the $l_j$'s correspond to distinct values of $i_X(x)$ except for $l_1=l_{\min}$, which may or may not be an information value.  Consider a subinterval $[l_k, l_{k+1}]$, $1\leq k \leq s-1$, which is illustrated in Figure \ref{fig: geogebra stairs}.
We will now derive a lower bound on $||F_{L_{n}} - F_{I_{X}} ||_{1}$ by integrating the area between the information CDF and the leakage CDF within this subinterval. Then, we select the particular subinterval that involves a type $P_n$ with the worst exponent, which maximizes the lower bound.}

We denote the value of $F_{I_{X}}(l)$ with $l_k \leq l < l_{k+1}$ as $Q_{k}$. If the CDFs cross within this range, this occurs at $l=F_{L_{n}}^{-1}(Q_{k})$. We can write,
\begin{equation} \label{eq: Q1 def}
    Q_{k} = \quad \sum_{\mathclap{\substack{y^{n}\in\mathcal{Y}^{ n} : \\ l(y^n) < F_{L_{n}}^{-1} (Q_{k})}}} \quad Q_{Y^{n}}(y^{n}) + \;\; \alpha \;\sum_{\mathclap{\substack{y^{n}\in\mathcal{Y}^{ n} : \\ l(y^n) = F_{L_{n}}^{-1} (Q_{k})}}}\quad Q_{Y^{n}}(y^{n}),
\end{equation}
where $0 \leq \alpha \leq 1$ is a constant.
Consider the $L^{1}$ norm:
\begin{align}
    &\int^{l_{\max}}_{l_{\min}} \left| F_{L_{n}}(l) - F_{I_{X}}(l) \right| \mathrm{d}l \geq \int_{l_k}^{l_{k+1}} \left| F_{L_{n}}(l) - F_{I_{X}}(l) \right| \mathrm{d}l \nonumber \\
    &= \int_{l_k}^{l_{k+1}} \! \bigg( \sum_{y^{n}\in \mathcal{Y}^{n}} Q_{Y^{n}}(y^{n})\mathds{1}[l(y^n) \leq l] - Q_{k} \bigg)^{+} \mathrm{d}l + \int_{l_k}^{l_{k+1}} \bigg(Q_{k} - \!\! \sum_{y^{n}\in \mathcal{Y}^{n}} \! Q_{Y^{n}}(y^{n}) \mathds{1}[l(y^n) \leq l] \bigg)^{+} \mathrm{d}l, \label{eq: L1 rate lb proof l2}
\end{align}
where $(z)^{+}=\max\{z, 0\}$. Taking the first term:
\begin{align}
    & \int_{l_k}^{l_{k+1}} \bigg( \sum_{y^{n}\in \mathcal{Y}^{n}} Q_{Y^{n}}(y^{n})\mathds{1}[l(y^n) \leq l] - Q_{k} \bigg)^{+} \mathrm{d}l \nonumber\\
    &= \int_{l_k}^{l_{k+1}} \sum_{y^{n}\in \mathcal{Y}^{n}} Q_{Y^{n}}(y^{n}) \big( \mathds{1}[l(y^n)\leq l] - \mathds{1}\left[l(y^n)< F_{L_{n}}^{-1}(Q_{k})\right] - \alpha \: \mathds{1}\left[l(y^n)=F_{L_{n}}^{-1}(Q_{k}) \right] \big)^{+} \mathrm{d}l \label{eq: L1 rate lb proof l3}\\
    &\geq \int_{l_k}^{l_{k+1}} \sum_{y^{n}\in \widetilde{\mathcal{Y}}^{n}} Q_{Y^{n}}(y^{n}) \big( \mathds{1}[l(y^n)\leq l] - \mathds{1}\left[l(y^n)< F_{L_{n}}^{-1}(Q_{k})\right] - \alpha \: \mathds{1}\left[l(y^n)=F_{L_{n}}^{-1}(Q_{k}) \right] \big)^{+} \mathrm{d}l \label{eq: L1 rate lb proof l4}\\
    &= \int_{l_k}^{l_{k+1}}  \sum_{y^{n}\in \widetilde{\mathcal{Y}}^{n}} Q_{Y^{n}}(y^{n}) \mathds{1}\left[ F_{L_{n}}^{-1}(Q_{k}) < l(y^n) \leq l \right] \mathrm{d}l \nonumber \\ &\qquad\qquad\qquad\qquad\qquad + \int_{l_k}^{l_{k+1}}  \sum_{y^{n}\in \widetilde{\mathcal{Y}}^{n}} Q_{Y^{n}}(y^{n}) (1-\alpha)\mathds{1}\left[ F_{L_{n}}^{-1}(Q_{k}) = l(y^n) \leq l \right]  \mathrm{d}l, \label{eq: L1 norm lb proof 1}
\end{align}
where (\ref{eq: L1 rate lb proof l3}) applies (\ref{eq: Q1 def}), and (\ref{eq: L1 rate lb proof l4}) removes terms for which $y^{n} \notin \widetilde{\mathcal{Y}}^{n}$. Lastly, (\ref{eq: L1 norm lb proof 1}) rearranges the indicator terms. To understand the last step, note that the sum of the three indicator terms in (\ref{eq: L1 rate lb proof l4}) is either $1$ if $ F_{L_{n}}^{-1}(Q_{k}) < l(y^n) \leq l $, $(1 - \alpha)$ if $ F_{L_{n}}^{-1}(Q_{k}) = l(y^n) \leq l$, or else non-positive.
We proceed with the first term of (\ref{eq: L1 norm lb proof 1}), and say that for $n$ large enough,
\begin{align}
    \int_{l_k}^{l_{k+1}} & \sum_{y^{n}\in \widetilde{\mathcal{Y}}^{n}} Q_{Y^{n}}(y^{n}) \mathds{1}\left[ F_{L_{n}}^{-1}(Q_{k}) < l(y^n) \leq l \right] \mathrm{d}l \nonumber \\
    &\geq \int_{l_k}^{l_{k+1}} \sum_{y^{n}\in  \widetilde{\mathcal{Y}}^{n}} Q_{Y^{n}}(y^{n}) \mathds{1}\left[ F_{L_{n}}^{-1}(Q_{k}) < l(y^n) \right] \label{eq: L1 rate lb l5} \mathds{1}\left[l \geq i_{X}(x^{\star}(y^{n})) - \textcolor{black}{\theta_{2}(n)} 2^{-nK(P_{n})} \right] \mathrm{d}l \\
    &\geq \sum_{y^{n}\in S^{1}_{l_{k+1}}} Q_{Y^n}(y^{n}) \int^{l_{k+1}}_{l_k} \mathds{1}\left[l \geq i_{X}(x^{\star}(y^{n})) - \textcolor{black}{\theta_2(n)} 2^{-nK(P_{n})} \right]\mathrm{d}l \label{eq: L1 rate lb proof l6} \\
    &= \sum_{y^{n} \in S^{1}_{l_{k+1}}} Q_{Y^{n}}(y^{n}) \min\left\{ \textcolor{black}{\theta_{2}(n)} 2^{-nK(P_{n})}, l_{k+1}-l_k \right\}, \label{eq: lb terms 1}
\end{align} 
where 
\begin{equation*}
  S^{1}_{l_{k+1}} \defeq \left\{
     y^{n} \in  \widetilde{\mathcal{Y}}^{n}: 
       i_{X}(x^{\star}(y^{n}))=l_{k+1} \; , \;  l(y^n) > F_{L_{n}}^{-1}(Q_{k})
  \right\},
\end{equation*}
(\ref{eq: L1 rate lb l5}) applies Lemma \ref{lemma: part 1} and (\ref{eq: L1 rate lb proof l6}) removes $y^n$ terms for which $i_{X}(x^\star (y^n)) \neq l_{k+1}$. The second term in (\ref{eq: L1 norm lb proof 1}) can be treated identically, giving
\begin{align}
    \int_{l_k}^{l_{k+1}}  \sum_{y^{n}\in \widetilde{\mathcal{Y}}^{n}} Q_{Y^{n}}(y^{n}) \alpha \mathds{1}&\left[l < l(y^n) = F_{L_{n}}^{-1}(Q_{k}) \right] \mathrm{d}l \nonumber \\& \geq (1-\alpha) \!\! \sum_{y^{n} \in S^{2}_{l_{k+1}}} Q_{Y^{n}}(y^{n}) \min\left\{ \textcolor{black}{\theta_2(n)} 2^{-nK(P_{n})}, l_{k+1}-l_k \right\}, \label{eq: lb terms 2}
\end{align}
where
\begin{equation*}
  S^{2}_{l_{k+1}} \defeq \left\{
     y^{n} \in \widetilde{\mathcal{Y}}^{n}: 
        i_{X}(x^{\star}(y^{n}))=l_{k+1} \;,\; l(y^n) = F_{L_{n}}^{-1}(Q_{k})
  \right\}.
\end{equation*}
Returning to (\ref{eq: L1 rate lb proof l2}), we now take the second term.
\begin{align}
    & \int_{l_k}^{l_{k+1}} \bigg(Q_{k} - \sum_{y^{n}\in \mathcal{Y}^{n}} Q_{Y^{n}}(y^{n}) \mathds{1}[l(y^n) \leq l] \bigg)^{+} \mathrm{d}l \nonumber \\
    &=\int_{l_k}^{l_{k+1}} \!\! \sum_{y^{n} \in \mathcal{Y}^{n}} Q_{Y^{n}}(y^{n}) \big(- \mathds{1}[l(y^n)\leq l] + \mathds{1}\left[l(y^n)< F_{L_{n}}^{-1}(Q_{k})\right] \label{eq: L1 rate proof lb l9} + \alpha\mathds{1}\left[ l(y^n)= F_{L_{n}}^{-1}(Q_{k}) \right]\big)^{+} \mathrm{d}l  \\
    &\geq \int_{l_k}^{l_{k+1}} \!\! \sum_{y^{n} \in \widetilde{\mathcal{Y}}^{n}} Q_{Y^{n}}(y^{n}) \big(- \mathds{1}[l(y^n)\leq l] + \mathds{1}\left[l(y^n)< F_{L_{n}}^{-1}(Q_{k})\right] \label{eq: L1 rate proof lb l10} + \alpha\mathds{1}\left[ l(y^n)= F_{L_{n}}^{-1}(Q_{k}) \right]\big)^{+} \mathrm{d}l  \\
    &= \int_{l_k}^{l_{k+1}}  \sum_{y^{n}\in \widetilde{\mathcal{Y}}^{n}} Q_{Y^{n}}(y^{n}) \mathds{1}\left[l < l(y^n) < F_{L_{n}}^{-1}(Q_{k}) \right]  \mathrm{d}l \nonumber \\ & \qquad\qquad\qquad\qquad\qquad\qquad+ \int_{l_k}^{l_{k+1}}  \sum_{y^{n}\in \widetilde{\mathcal{Y}}^{n}} Q_{Y^{n}}(y^{n}) \alpha \mathds{1}\left[l < l(y^n) = F_{L_{n}}^{-1}(Q_{k}) \right] \mathrm{d}l \label{eq: L1 norm lb proof 2}
\end{align}
where (\ref{eq: L1 rate proof lb l9}) applies (\ref{eq: Q1 def}), (\ref{eq: L1 rate proof lb l10}) removes terms for which $y^{n} \notin \widetilde{\mathcal{Y}}^{n}$, and (\ref{eq: L1 norm lb proof 2}) rearranges indicator terms. 
To understand the last step, note that the sum of the three indicator terms in (\ref{eq: L1 rate proof lb l10}) is either $1$ if $l < l(y^n) < F_{L_{n}}^{-1}(Q_{k})$, $\alpha$ if $l < l(y^n) = F_{L_{n}}^{-1}(Q_{k})$, or else non-positive.
We proceed with the first term of (\ref{eq: L1 norm lb proof 2}). If $n$ is large enough,
\begin{align}
    \int_{l_k}^{l_{k+1}}  &\sum_{y^{n}\in \widetilde{\mathcal{Y}}^{n}} Q_{Y^{n}}(y^{n}) \mathds{1}\left[l < l(y^n) < F_{L_{n}}^{-1}(Q_{k}) \right]  \mathrm{d}l \nonumber \\
    &\geq \int_{l_k}^{l_{k+1}} \sum_{y^{n}\in \widetilde{\mathcal{Y}}^{n}} Q_{Y^{n}}(y^{n})\mathds{1}\left[l(y^n) < F_{L_{n}}^{-1}(Q_{k}) \right] \mathds{1}\left[l < i_{X}(x^{\star}(y^{n}))- \textcolor{black}{\theta_1(n)} 2^{-nK(P_{n})} \right] \mathrm{d}l \label{eq: L1 rate lb proof l11}\\
    &\geq \sum_{y^{n}\in S^{3}_{l_{k+1}}} Q_{Y^{n}}(y^{n}) \int^{l_{k+1}}_{l_k} \mathds{1}\left[l < i_{X}(x^{\star}(y^{n})) - \textcolor{black}{\theta_1(n)} 2^{-nK(P_{n})} \right]\mathrm{d}l \label{eq: L1 rate lb proof l12} \\
    &= \sum_{y^{n}\in S^{3}_{l_{k+1}}} Q_{Y^{n}}(y^{n}) \left( l_{k+1} - l_k - \textcolor{black}{\theta_1(n)} 2^{-nK(P_{n})} \right)^{+}, \label{eq: lb terms 3}
\end{align}
where 
\begin{equation*}
  S^{3}_{l_{k+1}} \defeq \left\{
     y^{n} \in \widetilde{\mathcal{Y}}^{n}: 
    i_{X}(x^{\star}(y^{n}))=l_{k+1} \;,\; l(y^n) < F_{L_{n}}^{-1}(Q_{k})
  \right\},
\end{equation*}
(\ref{eq: L1 rate lb proof l11}) applies Lemma \ref{lemma: part 1} and (\ref{eq: L1 rate lb proof l12}) removes $y^n$ terms for which $i_{X}(x^\star (y^n)) \neq l_{k+1}$. The second term in (\ref{eq: L1 norm lb proof 2}) can be treated identically, giving 
\begin{align}
    \int_{l_k}^{l_{k+1}} \sum_{y^{n}\in \widetilde{ \mathcal{Y}}^{n} } Q_{Y^{n}}(y^{n}) \alpha \mathds{1}&\left[l < l(y^n) = F_{L_{n}}^{-1}(Q_{k}) \right] \mathrm{d}l \nonumber \\
    &\geq \alpha  \sum_{y^{n}\in S^{2}_{l_{k+1}}} Q_{Y^{n}}(y^{n}) \left( l_{k+1} - l_k - \textcolor{black}{\theta_1(n)} 2^{-nK(P_{n})} \right)^{+} . \label{eq: lb terms 4}
\end{align}
Finally combining (\ref{eq: lb terms 1}), (\ref{eq: lb terms 2}), (\ref{eq: lb terms 3}) and (\ref{eq: lb terms 4})  we have:
\begin{align}
    &\int_{l_{\min}}^{l_{\max}} \left| F_{L_{n}}(l) - F_{I_{X}}(l) \right| \mathrm{d}l \nonumber \\
    &\geq \! \sum_{y^{n}\in S^{1}_{l_{k+1}}} \!\!\!\! Q_{Y^{n}}(y^{n}) \min \! \left\{ \textcolor{black}{\theta_2(n)} 2^{-nK(P_{n})},l_{k+1}-l_k \right\} \! + \!\!\!\! \sum_{y^{n} \in S^{3}_{l_{k+1}}} \!\!\!\! Q_{Y^{n}}(y^{n}) \! \left( l_{k+1}\! -\! l_k \!- \! \textcolor{black}{\theta_1(n)} 2^{-nK(P_{n})} \right)^{+}\nonumber \\&\qquad + \sum_{y^{n}\in S_{l_{k+1}}^{2}} Q_{Y^{n}}(y^{n}) \min\left\{ \textcolor{black}{\theta_2(n)} 2^{-nK(P_{n})}, \left( l_{k+1} - l_k - \textcolor{black}{\theta_1(n)}2^{-nK(P_{n})} \right)^{+} \right\} \nonumber \\
    &\geq \sum_{y^{n}\in S_{l_{k+1}}} Q_{Y^{n}}(y^{n}) \min\left\{ \textcolor{black}{\theta_2(n)} 2^{-nK(P_{n})} , \left( l_{k+1} - l_k - \textcolor{black}{\theta_1(n)}2^{-nK(P_{n})} \right)^{+} \right\} \label{eq: L1 rate lb proof l17} \\
    &= \textcolor{black}{\theta_2(n)} \sum_{y^{n}\in S_{l_{k+1}} } Q_{Y^{n}}(y^{n}) 2^{-nK(P_{n})}, \label{eq: L1 rate lb proof l18}
\end{align} 
where $S_{l_{k+1}}$ is the union of $S^{1}_{l_{k+1}}$, $S^{2}_{l_{k+1}}$ and $S^{3}_{l_{k+1}}$, i.e., the set of all sequences $y^{n} \in \widetilde{\mathcal{Y}}^n$ such that $i_{X}(x^{\star}(y^{n}))=l_{k+1}$. We recognise that, for all $y^{n} \in \widetilde{\mathcal{Y}}^{n}$, $\textcolor{black}{\theta_2(n)}2^{-nK(P_{n})}$ and $\textcolor{black}{\theta_1(n)}2^{-nK(P_{n})}$ can be made arbitrarily small. In (\ref{eq: L1 rate lb proof l18}) we take $n$ to be large enough that $\textcolor{black}{\theta_2(n)}2^{-nK(P_{n})}$ is always the minimiser of (\ref{eq: L1 rate lb proof l17}). 
We next define the domain $D_{l_{k+1}}$ as follows
\begin{equation*}
    D_{l_{k+1}} \defeq \bigcup_{x:i_X(x)=l_{k+1}}D_x,
\end{equation*}
which means that $P_n \in \widetilde{\mathcal{P}}_n \cap D_{l_{k+1}} \Rightarrow P_n \in S_{l_{k+1}}$ by Remark \ref{remark: x1 = x*}.
Continuing directly from (\ref{eq: L1 rate lb proof l18}) and applying the law of total probability:
\begin{align}
    \int_{l_{\min}}^{l_{\max}} &\left| F_{L_{n}}(l) - F_{I_{X}}(l) \right| \mathrm{d}l \nonumber \\
    &\geq \textcolor{black}{\theta_2(n)} \sum_{P_{n}\in \widetilde{\mathcal{P}}_{n} \cap {D}_{l_{k+1}}}\sum_{x'} Q_{Y^{n}|X=x'}(T(P_{n}))Q_{X}(x') 2^{-n K(P_{n})} \label{eq: same steps applied} \\
    &\geq \frac{\textcolor{black}{\theta_2(n)} \min_{x}Q_{X}(x)}{(n+1)^{|\mathcal{Y}|}} \sum_{P_{n}\in \widetilde{\mathcal{P}}_{n} \cap {D}_{l_{k+1}}}\sum_{x'}2^{-n D(P_{n}||Q_{x'})} 2^{-n D(P_{n}||Q_{x_{2}(P_{n})})}2^{n D(P_{n}||Q_{x^{\star}})} \nonumber \\
    &\geq \frac{\textcolor{black}{\theta_2(n)} \min_{x}Q_{X}(x)}{(n+1)^{|\mathcal{Y}|}} \sum_{P_{n}\in \widetilde{\mathcal{P}}_{n} \cap {D}_{l_{k+1}}} 2^{-nD(P_{n}||Q_{x_{2}(P_{n})})} \label{eq: th 2 lb 118} \\
    &\geq \frac{\textcolor{black}{\theta_2(n)} \min_{x}Q_{X}(x)}{(n+1)^{|\mathcal{Y}|}} 2^{-n \inf_{P_{n}\in \widetilde{\mathcal{P}}_{n} \cap {D}_{l_{k+1}}}D(P_{n}||Q_{x_{2}(P_{n})})}, \label{eq: L1 rate lb last line}
\end{align}
where in (\ref{eq: th 2 lb 118}) we take only $x' = x^\star$ terms.
Consider the exponent. Firstly, note that $l_{k+1}$ was arbitrary. We can therefore select $l_{k+1}$ such that the argument of $\inf_{P_{n}\in \widetilde{\mathcal{P}}_{n} } D(P_{n}||Q_{x_{2}(P_{n})})$, is in $D_{l_{k+1}}$, i.e.,
\begin{equation*}
    \inf_{P_{n}\in \widetilde{\mathcal{P}}_{n} \cap {D}_{l_{k+1}}}D(P_{n}||Q_{x_{2}(P_{n})}) = \inf_{P_{n}\in \widetilde{\mathcal{P}}_{n} } D(P_{n}||Q_{x_{2}(P_{n})}).
\end{equation*}
This, combined with (\ref{eq: L1 rate lb last line}) \textcolor{black}{and the fact that $\lim_{n \to \infty} \frac{1}{n} \log \frac{\theta_2(n) \min_{x}Q_{X}(x)}{(n+1)^{|\mathcal{Y}|}}=0$} gives
\begin{align} \label{eq: th 2 lb with lim}
    \lim_{n \to \infty} \frac{1}{n} \log ||F_{L_{n}} - F_{I_{X}} ||_{1} &\geq - \lim_{n \to \infty} \inf_{P_{n}\in \widetilde{\mathcal{P}}_{n}} D(P_{n}||Q_{x_{2}(P_{n})})\\
    & \textcolor{black}{ = -\lim_{n \to \infty} \inf_{P_{n}:D(P_{n}||Q_{x_{1}(P_{n})}) \leq C_{n}- \frac{1}{\sqrt{n}}} D(P_{n}||Q_{x_{2}(P_{n})}).}\label{eq: th 2 lb with constraint}
\end{align}
{\color{black}To argue \eqref{eq: th 2 lb with constraint}, let us first observe that
\begin{align*}
\inf_{P:D(P||Q_{x_{1}(P)}) \leq C} D(P||Q_{x_{2}(P)}) = C.
\end{align*}
which easily follows from the fact that if $\hat{P}$ is the argument of the minimum Chernoff information $C$, then $D(\hat{P}||Q_{x_{1}(\hat{P})}) = D(\hat{P}||Q_{x_{2}(\hat{P})}) = C$.
As $n \to \infty$, by Remark~\ref{remark: chernoff}, the constraint in \eqref{eq: th 2 lb with constraint} approaches the limit $C$, i.e., $C_n - \frac{1}{\sqrt{n}} \to C$, which means that every empirical distribution satisfying the constraint $ D(P||Q_{x_1(P)}) < C$ is eventually considered in the constraint set of an optimization problem corresponding to some $n$. Therefore, as the empirical distributions are dense in the probability simplex, we can consider a sequence of them $P_k \to \hat{P}$ (in $L^{1}$-norm) in the set $D(P||Q_{x_1(P)}) < C$ with $x_2(P_k) = x_2(\hat{P})$ for $k\geq 1$. The latter condition is satisfied for arbitrarily close distributions to $\hat{P}$ as KL divergence is a continuous function and $x_2(P_k)$ is a discrete minimiser of it.
Therefore, the limit in (\ref{eq: th 2 lb with constraint}) can be bounded as follows
\begin{align*}
    \lim_{n \to \infty} \inf_{P_{n}:D(P_{n}||Q_{x_{1}(P_{n})}) \leq C_{n}- \frac{1}{\sqrt{n}}} D(P_{n}||Q_{x_{2}(P_{n})})  & \leq \lim_{P_k \to \hat{P}} D(P_k || Q_{x_2(P_k)}) \nonumber\\
    &= \lim_{P_k \to \hat{P}} D(P_k || Q_{x_2(\hat{P})}) \nonumber\\
    &= D(\hat{P} ||Q_{x_2(\hat{P})})\nonumber\\& = C,
\end{align*}
where the second equality follows from standard continuity arguments for KL divergence \cite[Sec. III-A]{KLcontinuous} and the fact that the distributions are discrete.}
This along with (\ref{eq: th 2 lb with constraint}) gives
\begin{align*}
   \lim_{n \to \infty} \frac{1}{n} \log ||F_{L_{n}} - F_{I_{X}} ||_{1} \geq -C,
\end{align*}
for information theoretic leakage functions with the derivative property.
This, combined with (\ref{eq: th 2 general}) proves the equality part of Theorem \ref{theorem: rate of L1 convergence}.

\subsection{Global Leakage Proofs}
\label{section: global leakage proofs}

The next statement is analogous to Lemma \ref{lemma: part 1}, \textcolor{black}{but considers the function $h(P,Q) = g_1(f(P,Q))$ rather than $f(P,Q)$.}
\begin{lemma} \label{Lemma: global part 1}
    \textcolor{black}{Let $g_1$ be a function satisfying Condition~\ref{assumption: g2}, and let $h(P,Q) = g_1(f(P,Q))$ with $f$ being a pointwise leakage function.
    }
    For any observation $y^{n}$ whose type $P_{n} \in \widetilde{\mathcal{P}}^{n}$, $h(Q_{X|Y^{n}=y^{n}},Q_{X})$ is bounded by $h(E_{x^{\star}},Q_{X})$ as follows for large enough $n$.
    \textcolor{black}{If $f$ satisfies Condition \ref{assumption: f double bound}},
    \textcolor{black}{
    \begin{equation} \label{eq: Lemma 6 cond1}
        h(E_{x^{\star}},Q_{X}) -\theta_1'(n) 2^{-nK(P_{n})} \leq h(Q_{X|Y^{n}=y^{n}},Q_{X}),
    \end{equation}
    and if $f$ satisfies Condition \ref{assumption: f double bound} with $A>0$,
    }
    \begin{equation} \label{eq: lemma 6 cond2}
        h(E_{x^{\star}},Q_{X}) - \textcolor{black}{\theta_1'(n)} 2^{-nK(P_{n})} \leq h(Q_{X|Y^{n}=y^{n}},Q_{X}) \leq h(E_{x^{\star}},Q_{X}) - \textcolor{black}{\theta_2'(n)} 2^{-nK(P_{n})}
    \end{equation}
    where
    \textcolor{black}{
    \begin{align*}
        \theta'_{1}(n) &\defeq B'' n^{|b|} \left( 1 + O\left( \frac{1}{\sqrt{n}} \right) \right),  \\
        \theta'_{2}(n) &\defeq A'' n^{-|a|} \left( 1 - O\left( \frac{1}{\sqrt{n}} \right) \right), 
    \end{align*}
    }
     and \textcolor{black}{$B'', A'' >0$} are finite constants.
     \begin{proof}
        \textcolor{black}{
        Let us first argue that if Condition~\ref{assumption: f double bound} holds for $f$, then it also holds for $h$. For a pointwise leakage function $f$ satisfying Condition~\ref{assumption: f double bound}, there exist neighbourhoods around the extreme points of the probability simplex, and constants $A \geq 0$, $B>0$ and $a, b \in \mathbb{R}$ such that
    \begin{equation}\label{eq: h condition 1 only}
        f(E_i, Q) - B(1 - p_i) \left( \log \frac{1}{1- p_i}\right)^{b} \leq  f(P,Q) \leq f(E_i, Q) - A(1 - p_i) \left( \log \frac{1}{1- p_i}\right)^{a}
    \end{equation}
    if $P=(p_1, \ldots, p_i, \ldots, p_{|\mathcal{X}|}) \in \mathcal{P}_{\mathcal{X}}$ lies in the corresponding neighbourhood of $E_i$ for $i \in \{1, \dots, |\mathcal{X}| \}$. As $g_1$ is a monotonically increasing function, \eqref{eq: h condition 1 only} becomes
        \begin{align}
            g_1\left( f(E_i, Q) - B(1 - p_i) \left( \log \frac{1}{1- p_i}\right)^{b} \right) &\leq h(P,Q) \nonumber\\ &\leq g_1\left( f(E_i, Q) - A(1 - p_i) \left( \log \frac{1}{1- p_i}\right)^{a} \right) \label{eq: use g1 monotonic}
        \end{align} 
 It follows from the differentiability of $g_1$ and the strict positivity of the derivative $g_1'>0$ that $\frac{g_1'(u)}{2}\delta \leq {g_1(u)- g_1(u-\delta)} \leq \frac{3g_1'(u)}{2}\delta$ for $0\leq  \delta < \delta_u$, where $\delta_u$ is sufficiently small. 
 Note that $ \lim_{p \to 1^-}(1 - p) \left( \log \frac{1}{1- p}\right)^{s}=0$ for any $s \in \mathbb{R}$.
 Thus, 
 within the neighbourhoods of extreme points identified in Condition \ref{assumption: f double bound},
 we consider smaller neighbourhoods that contain $P$'s such that $\max\{A, B\}(1 - p_i) \left( \log \frac{1}{1- p_i}\right) \leq \delta_0$ for all $i$, where $\delta_0= \min\{\delta_u : u=f(E_i, Q), i \in \{1, \dots, |\mathcal{X}|\} \}$. 
By combining this with the above condition on $g$ at $u=f(E_i, Q)$, we have
        \begin{align}
            h(E_i, Q) - \frac{3g_{1, \max}'}{2}B (1 - p_i) \left( \log \frac{1}{1- p_i}\right)^{b} &\leq h(P,Q)\nonumber \\ & \leq h(E_i,Q) - \frac{g_{1, \min}'}{2}A (1 - p_i) \left( \log \frac{1}{1- p_i}\right)^{a}, \label{eq: use taylors h lemma}
        \end{align}
        for $i \in \{1, \dots, |\mathcal{X}|\}$, where $g_{1, \max}'= \max_i g_{1}'(f(E_i, Q)) >0$ and $g_{1, \min}'= \min_i g_{1}'(f(E_i, Q))>0$. This proves that $h$ satisfies Condition~\ref{assumption: f double bound}. Now we can apply Lemma~\ref{lemma: part 1} for $h$ with $\frac{3g_{1, \max}'}{2}B$ and $\frac{g_{1, \min}'}{2}A$ replacing $B$ and $A$, yielding \eqref{eq: Lemma 6 cond1} and \eqref{eq: lemma 6 cond2} with $B'' = \frac{3g_{1, \max}'}{2} B'$ and $A'' = \frac{g_{1, \min}'}{2} A'$.
        }
     \end{proof}
\end{lemma}
\subsubsection{Proof of Theorem \ref{theorem: global rate}} \label{section: proof of th 3}
We use \textcolor{black}{(\ref{eq: Lemma 6 cond1})} of Lemma \ref{Lemma: global part 1}  to prove the upper bound of $\mathcal{L}_\infty - \mathcal{L}_n$. 
\textcolor{black}{Note that this applies for $f$ satisfying Condition \ref{assumption: f double bound} either with or without the restriction $A>0$.}
Many of the steps are are similar to the proof of Lemma \ref{lemma: part 3 new}. First, we consider $\mathbb{E}_{Y^n} [g_1 (l(Y^n))]$ :
\begin{align}
    &\sum_{y^{n}\in \mathcal{Y}^{n}} Q_{Y^{n}}(y^{n}) h(Q_{X|Y^{n}=y^{n}},Q_{X}) \nonumber \\
    &= \sum_{y^{n}\in \widetilde{\mathcal{Y}}^{n}} Q_{Y^{n}}(y^{n}) h(Q_{X|Y^{n}=y^{n}},Q_{X}) + \sum_{y^{n} \notin \widetilde{\mathcal{Y}}^{n}} Q_{Y^{n}}(y^{n}) h(Q_{X|Y^{n}=y^{n}},Q_{X}) \nonumber \\
    &\geq \sum_{y^{n}\in \widetilde{\mathcal{Y}}^{n}} Q_{Y^{n}}(y^{n}) \left( h(E_{x^{\star}}, Q_{X}) - \textcolor{black}{\theta'_1(n)} 2^{-nK(P_{n})} \right) - \max_{P \in \mathcal{P}_{\mathcal{X}}} |h(P,Q_X)| \sum_{y^{n} \notin \widetilde{\mathcal{Y}}^{n}} Q_{Y^{n}}(y^{n}) \label{eq: global lb l2} \\
    &\geq \sum_{y^{n}\in \widetilde{\mathcal{Y}}^{n}} Q_{Y^{n}}(y^{n}) \left( h(E_{x^{\star}}, Q_{X}) - \textcolor{black}{\theta'_1(n)} 2^{-nK(P_{n})} \right) - \max_{P \in \mathcal{P}_{\mathcal{X}}} |h(P,Q_X)| \psi \label{eq: global lb l3} \\
    &= \sum_{x \in \mathcal{X}}Q_{Y^{n}}(S_{x})h(E_{x},Q_{X})  -\sum_{y^n\notin \tilde{\mathcal{Y}}^n}Q_{Y^n}(y^n) h(E_{x^*}, Q_X) \nonumber \\ & \qquad\qquad\qquad\qquad\qquad\qquad - \textcolor{black}{\theta'_1(n)}  \sum_{x \in \mathcal{X}}\sum_{P_{n} \in \widetilde{\mathcal{P}}_{n} \cap D_{x}} Q_{Y^{n}|X=x} (T(P_{n})) Q_{X}(x) 2^{-nK(P_{n})} - \psi' \label{eq: global lb l4} \\
    &\geq \sum_{x \in \mathcal{X}}Q_{Y^{n}}(S_{x})h(E_{x},Q_{X}) - \textcolor{black}{\theta'_1(n)} \sum_{x \in \mathcal{X}}\sum_{P_{n} \in \widetilde{\mathcal{P}}_{n} \cap D_{x}} Q_{Y^{n}|X=x} (T(P_{n})) Q_{X}(x) 2^{-nK(P_{n})} - 2\psi' \nonumber \\
    &\geq \sum_{x \in \mathcal{X}}\left(Q_X(x)h(E_x, Q_X)-\phi |h(E_x, Q_X)|\right) - \textcolor{black}{\theta'_1(n)}  \sum_{x \in \mathcal{X}}\sum_{P_{n} \in \widetilde{\mathcal{P}}_{n} \cap D_{x}} 2^{-nD(P_{n}||Q_{x})} 2^{-nK(P_{n})} - 2\psi' \label{eq: global lb l4.5} \\
    &\geq \sum_{x \in \mathcal{X}}\! Q_{X}(x) h(E_x, Q_X) \! - \! |\mathcal{X}| \phi \max_x | h(E_{x},Q_{X})| \!- \! \textcolor{black}{\theta'_1(n)}  \! \sum_{x \in \mathcal{X}}\sum_{P_{n} \in \widetilde{\mathcal{P}}_{n} \cap D_{x}} \!\!\!\! 2^{-nD(P_{n}||Q_{x})} 2^{-nK(P_{n})} \!- \! 2\psi' \label{eq: global lb l5}
\end{align}
where $\psi' \defeq \max_{P \in \mathcal{P}_{\mathcal{X}}} |h(P,Q_X)| \psi$, (\ref{eq: global lb l2}) applies Lemma \ref{Lemma: global part 1} \textcolor{black}{and the boundedness of $h(P,Q_X)$ for  $P \in \mathcal{P}_{\mathcal{X}}$, which follows from the convexity of $h$ in its first argument and the compactness of $\mathcal{P}_{\mathcal{X}}$}, (\ref{eq: global lb l3}) applies Lemma \ref{lemma: not tilde prob}, (\ref{eq: global lb l3}) applies Lemma \ref{lemma: not tilde prob}, and (\ref{eq: global lb l4.5}) applies Lemma \ref{lemma: part 2}. Consider the third term.
\begin{align}
    \textcolor{black}{\theta'_1(n)}  \sum_{x \in \mathcal{X}}\sum_{P_{n} \in \widetilde{\mathcal{P}}_{n} \cap D_{x}}& 2^{-nD(P_{n}||Q_{x})} 2^{-nK(P_{n})} \nonumber \\
    &= \textcolor{black}{\theta'_1(n)}  \sum_{x \in \mathcal{X}}\sum_{P_{n} \in \widetilde{\mathcal{P}}_{n} \cap D_{x}} 2^{-nD(P_{n}||Q_{x})}2^{-nD(P_{n}||Q_{x_{2}(P_{n})})}2^{nD(P_{n}||Q_{x^{\star}})} \nonumber \\
    &\leq \textcolor{black}{\theta'_1(n)}  |\mathcal{X}| \sum_{P_{n} \in \widetilde{\mathcal{P}}_{n}} 2^{-nD(P_{n}||Q_{x_{2}(P_{n})})} \label{eq: global lb l6}\\
    & \leq \textcolor{black}{\theta'_1(n)}  |\mathcal{X}| (n+1)^{|\mathcal{Y}|}  2^{-n C_n}. \nonumber
\end{align}
In (\ref{eq: global lb l6}) we recognise that $2^{-nD(P_{n}||Q_{x})} \leq 2^{-nD(P_{n}||Q_{x^{\star}})}$ for any $x \in \mathcal{X}$. We can continue from (\ref{eq: global lb l5}).
\begin{align*}
    \sum_{y^{n}\in \mathcal{Y}^{n}} &Q_{Y^{n}}(y^{n}) h(Q_{X|Y^{n}=y^{n}},Q_{X}) \nonumber \\
    &\geq \sum_{x \in \mathcal{X}} Q_{X}(x) h(E_{x},Q_{X}) - |\mathcal{X}| \phi \max_{x}|h(E_{x},Q_{X})| - \textcolor{black}{\theta'_1(n)}  |\mathcal{X}| (n+1)^{|\mathcal{Y}|}  2^{-n C_n} - 2\psi'.
\end{align*}
\textcolor{black}{Now applying $g_{2}$, which satisfies Condition~\ref{assumption: g2} and is a strictly increasing function, we get}
\begin{align} 
    &\mathcal{L}_{n} = g_{2} \left( \sum_{y^n \in \mathcal{Y}^n} Q_{Y^n}(y^n) h(Q_{X|Y^n=y^n},Q_{X}) \right) \nonumber \\ 
    & \geq g_2 \left( \sum_{x \in \mathcal{X}} Q_{X}(x) h(E_{x},Q_{X}) - |\mathcal{X}| \phi \max_{x}|h(E_{x},Q_{X})| - \textcolor{black}{\theta'_1(n)}  |\mathcal{X}| (n+1)^{|\mathcal{Y}|}  2^{-n C_n} - 2\psi' \right) \nonumber \\
    &\textcolor{black}{\geq} g_{2} \!\! \left( \sum_{x \in \mathcal{X}} Q_{X}(x) h(E_{x},Q_{X}) \! \! \right)
    \!\! - \!\! \bigg( |\mathcal{X}|\phi \max_{x}|h(E_{x},Q_{X})| + \textcolor{black}{\theta'_1(n)}  |\mathcal{X}| (n+1)^{|\mathcal{Y}|}  2^{-n C_n}  \textcolor{black}{ + } 2\psi' \bigg) \textcolor{black}{\frac{3g'_{2}(u)}{2}}, \label{eq: global lb g taylor}
\end{align}
\textcolor{black}{where $u:= \sum_{x \in \mathcal{X}} Q_{X}(x) h(E_{x},Q_{X})$, \eqref{eq: global lb g taylor} assumes $n$ is large enough and uses the inequality $g_2(u)- g_2(u-\delta) \leq \frac{3g'(u)}{2}\delta$ for $0\leq  \delta < \delta_u$ with is sufficiently small $\delta_u$. The latter is a consequence of the differentiability and the strict positivity of the derivative of $g_2$.}  Finally, we can say
\begin{align} \label{eq: th 3 ub}
    \lim_{n \to \infty} \frac{1}{n} \log \left( \mathcal{L}_{\infty} - \mathcal{L}_n \right) \leq \lim_{n \to \infty} - C_n = -C,
\end{align}
where in the equality we have employed Remark \ref{remark: chernoff}.
We have the upper bound of Theorem \ref{theorem: global rate}.

Next, we find the lower bound of $\mathcal{L}_{\infty} - \mathcal{L}_n$ \textcolor{black}{for $f$ satisfying Condition \ref{assumption: f double bound} with $A>0$}. We will consider $y^{n} \in \widetilde{\mathcal{Y}}^{n}$ and $y^{n} \notin \widetilde{\mathcal{Y}}^{n}$ separately. Starting with the latter, we use the convexity of $h(P,Q)$ in $P$ to say
\begin{align*}
    h(Q_{X|Y^{n}=y^{n}},Q_{X}) &\leq \sum_{x \in \mathcal{X}} Q_{X|Y^{n}=y^{n}}(x) h(E_{x},Q_{X}),
\end{align*}
and
\begin{align}
    \sum_{y^{n} \notin \widetilde{\mathcal{Y}}^{n}} Q_{Y^{n}}(y^{n}) h(Q_{X|Y^{n}=y^{n}},Q_{X}) &\leq \sum_{x \in \mathcal{X}}Q_{X}(x) h(E_{x},Q_{X}) \sum_{y^{n} \notin \widetilde{\mathcal{Y}}^{n}} Q_{Y^{n}|X=x}(y^{n}), \label{eq: global ub bad set}
\end{align}
where the last line takes an average over $y^{n} \notin \widetilde{\mathcal{Y}}^{n}$ and makes use of Bayes' theorem on the RHS. 
Moving onto  $y^{n} \in \widetilde{\mathcal{Y}}^{n}$, 
we \textcolor{black}{first apply \eqref{eq: lemma 6 cond2} of Lemma \ref{Lemma: global part 1} }.
We then scale each term by $Q_{X|Y^{n}=y^{n}}(x)$ and sum over all $x \in 
\mathcal{X}$, giving
\begin{align}
    &\sum_{y^n \in \widetilde{ \mathcal{Y}}^n} Q_{Y^n} (y^n) h(Q_{X|Y^n=y^n},Q_X) \nonumber \\
    &\leq \sum_{x \in \mathcal{X}} Q_X(x) h(E_x,Q_X) \sum_{y^n \in \widetilde{ \mathcal{Y}}^n} Q_{Y^n|X=x}(y^n) - \textcolor{black}{\theta_2'(n)} \sum_{y^n \in \widetilde{ \mathcal{Y}}^n} Q_{Y^n}(y^n) 2^{-nK(P_n)} \nonumber \\
    & = \sum_{x \in \mathcal{X}} \! Q_X(x) h(E_x,Q_X) \!\! \sum_{y^n \in \widetilde{ \mathcal{Y}}^n} \!\! Q_{Y^n|X=x}(y^n) \!- \textcolor{black}{\theta_2'(n)} \!\! \sum_{P_n \in \widetilde{\mathcal{P}}_n} \sum_{x' \in \mathcal{X}} Q_{Y^n|X=x'}(T(P_n)) Q_X(x') 2^{-nK(P_n)} \nonumber \\
    &\leq \sum_{x \in \mathcal{X}} Q_X(x) h(E_x,Q_X) \sum_{y^n \in \widetilde{ \mathcal{Y}}^n} Q_{Y^n|X=x}(y^n) - \textcolor{black}{\theta_2'(n)} 2^{-n C_n} \label{eq: apply previously used steps},
\end{align}
where (\ref{eq: apply previously used steps}) applies the steps from (\ref{eq: same steps applied}-\ref{eq: L1 rate lb last line}). Summing (\ref{eq: global ub bad set}) and (\ref{eq: apply previously used steps}) gives
\begin{align}
    \sum_{y^n \in \mathcal{Y}^n} Q_{Y^n} (y^n) h(Q_{X|Y^n=y^n},Q_X) \leq \sum_{x \in \mathcal{X}} Q_X(x) h(E_x, Q_X) - \theta_2'(n) 2^{-nC_n}.
\end{align}
Finally, we can apply $g_2$ \textcolor{black}{and make use of Condition~2, which implies that $g_2(v)- g_2(v-\delta) \geq \frac{g'(v)}{2}\delta$ for $0\leq  \delta < \delta_v$ with is sufficiently small $\delta_v$ for any $v$ in the domain of $g_2$:}
\begin{align}
    \mathcal{L}_n &= g_2 \left( \sum_{y^n \in \mathcal{Y}^n} Q_{Y^n} (y^n) h(Q_{X|Y^n=y^n},Q_X) \right) \nonumber \\
    &\leq g_2 \left( \sum_{x \in \mathcal{X}} Q_X(x) h(E_x, Q_X) - \textcolor{black}{\theta_2'(n)} 2^{-nC_n} \right) \nonumber \\
    &\textcolor{black}{\leq} g_2 \left( \sum_{x \in \mathcal{X}} Q_X(x) h(E_x, Q_X) \right) -  \textcolor{black}{\frac{g_2'(v)}{2}\theta_2'(n)} 2^{-nC_n} \nonumber
\end{align}
where $v$ is $\sum_{x \in \mathcal{X}} Q_X(x) h(E_x, Q_X)$.
It follows that 
\begin{align*}
    \lim_{n \to \infty} \frac{1}{n} \log \left( \mathcal{L}_{\infty} - \mathcal{L}_{n} \right) &\geq  \lim_{n \to \infty} - C_n = - C,
\end{align*}
where the last equality makes use of Remark \ref{remark: chernoff}.
This, combined with (\ref{eq: th 3 ub}) completes the proof of Theorem \ref{theorem: global rate}.

{\color{black}
\subsection{Proof of Lemma~\ref{lemma: gradients}}\label{section: proof of gradient}}
        Consider the pair $(E_i, Q)$ for an $i \in \{1, 2, \ldots, |\mathcal{X}|\}$. By keeping $Q$ fixed, let us view $f$ as a function only in the first argument. As the function is differentiable at $E_i$, the gradient of $f$, $\nabla f$, is well defined and so are the directional derivatives. If the function satisfies the derivative property, then
        \begin{equation*}
            (U-E_{i}) \cdot \nabla f(P,Q) \big|_{P=E_{i}} < 0,
        \end{equation*}
        for all $U \in \mathcal{P}_{\mathcal{X}}$ such that $U \neq E_{i}$. \textcolor{black}{Defining $e_{ij}$ as the $j$th element of $E_i$} and rewriting this expression as a summation,
        \begin{align}
            (U-E_{i}) \cdot \nabla f(P,Q) \big|_{P=E_{i}} &= \sum_{j=1}^{|\mathcal{X}|} (u_{j}-e_{ij})\frac{\partial f(P,Q)}{\partial p_{j}}\bigg|_{P=E_{i}} \nonumber \\
            &= (u_{i}-1) \frac{\partial f(P,Q)}{\partial p_{i}}\bigg|_{P=E_{i}} + \sum_{j \neq i} u_{j} \frac{\partial f(P,Q)}{\partial p_{j}}\bigg|_{P=E_{i}} \nonumber \\
            &= - \sum_{j \neq i} u_{j} \frac{\partial f(P,Q)}{\partial p_{i}}\bigg|_{P=E_{i}} + \sum_{j \neq i} u_{j} \frac{\partial f(P,Q)}{\partial p_{j}}\bigg|_{P=E_{i}} \label{eq: cor2 l3} \\
            &= \sum_{j \neq i} u_{j} \left( \frac{\partial f(P,Q)}{\partial p_{j}} \bigg|_{P=E_{i}} -  \frac{\partial f(P,Q)}{\partial p_{i}} \bigg|_{P=E_{i}} \right) \label{eq: cor2 l4}\\
            &<0\nonumber
        \end{align}
        where (\ref{eq: cor2 l3}) follows from the fact that $U$ is a probability vector and thus $1 - u_{i} = \sum_{j \neq i} u_{j}$. {\color{black} For (\ref{eq: cor2 l4}) to be true for any $U \in \mathcal{P}_{\mathcal{X}}$ such that $U \neq E_{i}$, 
        it must be true for all $U = E_{j}$ where $j \neq i$. When $U = E_{j}$, carrying out the summation in (\ref{eq: cor2 l4}) leaves only the bracketed term corresponding to the index $j$. Thus, this term must be strictly negative for each $j \neq i$.
        For the other direction, consider \eqref{eq: cor2 l4}, which is 
        \begin{align*}
            (U-E_{i}) \cdot \nabla f(P,Q) \big|_{P=E_{i}}
            &= \sum_{j \neq i} u_{j} \left( \frac{\partial f(P,Q)}{\partial p_{j}} \bigg|_{P=E_{i}} -  \frac{\partial f(P,Q)}{\partial p_{i}} \bigg|_{P=E_{i}} \right).
        \end{align*}        
        Assume that the bracketed terms are strictly negative for $j \neq i$. As $U \neq E_i$, there is at least one component $u_j > 0$, $j \neq i$ with the rest of the components being non-negative. Therefore $(U-E_{i}) \cdot \nabla f(P,Q) \big|_{P=E_{i}}<0$, completing the proof of Lemma \ref{lemma: gradients}. 
}

{\color{black}
\subsection{Proof of Proposition~\ref{remark: differentials to conditions}} \label{section: proof of differentials to conds}
Fix a $Q \in \mathcal{P}^{\circ}_{\mathcal{X}}$, and note that if the function $f(\cdot, Q)$ is differentiable at an extreme point $E_i$, then we have that for every $\epsilon >0$, there exists a $\delta_{\epsilon}>0$ such that 
\begin{align}\label{eq:diff_grad}
    -\epsilon \leq \frac{f(P, Q)-f(E_i,Q) -(P-E_i)^{\intercal} \nabla f(E_i,Q)}{\left\lVert P-E_i \right\rVert_1}\leq \epsilon
\end{align}
for all $P$ around $E_i$ satisfying $\left\lVert P-E_i \right\rVert_1 \leq \delta_{\epsilon}$. Also, observe that if $P$ is a probability vector, then 
\begin{align}\label{eq:norm}
    \left\lVert P-E_i \right\rVert_1= \sum_{j=1}^{|\mathcal{X}|} | p_j - e_{ij} |= (1-p_i) + \sum_{j \neq i} p_j = 2 (1-p_i),
\end{align}
\textcolor{black}{where again, $e_{ij}$ is defined as the $j$th element of $E_i$.}

Now we prove the first statement that if the function $f$ is differentiable with respect to $P$ at all the extreme points $P=E_i$,  $i \in \{1, \dots, |\mathcal{X}| \}$, then $f$ satisfies Condition  \ref{assumption: f double bound} with $A=0$, $B>0$, and $b=0$. Let us fix an $\epsilon>0$ in  \eqref{eq:diff_grad} and consider a sufficiently small $\delta \leq \delta_{\epsilon}$. We make use of \eqref{eq:diff_grad}  and the strict local maximality of $f(E_i,Q)$ in the probability simplex $\mathcal{P}_{\mathcal{X}}$. We can write that for $i \in \{1, \dots, |\mathcal{X}| \}$ and all probability vectors $P \in \mathcal{P}_{\mathcal{X}}$ in the neighbourhood of $E_i$ where $\left\lVert P-E_i \right\rVert_1 \leq \delta$,
\begin{align}
    0 < f(E_i,Q) -  f(P,Q) &\leq -(P-E_i)^{\intercal} \nabla f(E_i,Q) + \epsilon \left\lVert P-E_i \right\rVert_1 \label{eq:diff_only_1}\\
      &\leq   | (P-E_i)^{\intercal} \nabla f(E_i,Q) | + \epsilon \left\lVert P-E_i \right\rVert_1\nonumber\\
      & \leq \lambda \left\lVert P-E_i \right\rVert_1 + \epsilon \left\lVert P-E_i \right\rVert_1\label{eq:diff_only_2}\\
      & = 2(\lambda + \epsilon)(1-p_i)\label{eq:diff_only_3}.
\end{align}
Here, the left inequality of \eqref{eq:diff_only_1} follows from Axiom \ref{item: A5} and the fact that $\delta$ is small enough that the $f(E_i,Q)$ is local strict maximum within the neighbourhood; the right inequality of \eqref{eq:diff_only_1} is due to \eqref{eq:diff_grad}; in \eqref{eq:diff_only_2}, $\lambda$ is the maximum magnitude of any element in $\nabla f (E_i, Q)$ over all $i$; and \eqref{eq:diff_only_3} applies \eqref{eq:norm}.  We have reached the first statement of Proposition~\ref{remark: differentials to conditions}, where Condition \ref{assumption: f double bound} is met with $A=0$, $B = 2(\lambda + \epsilon) >0$ and $b = 0$.}

{\color{black}Next, we consider the case when $f$ satisfies the derivative property. Recall from the proof of Lemma \ref{lemma: gradients} that we can write
\begin{align*}
    (P-E_i)^{\intercal} \nabla f(E_i,Q) &= \sum_{j \neq i} p_j \left( \frac{\partial f(P,Q)}{\partial p_j} \Bigg|_{P = E_i} - \frac{\partial f(P,Q)}{\partial p_i } \Bigg|_{P = E_i}\right).
\end{align*}
We know from Lemma \ref{lemma: gradients} that the bracketed term in the last line is strictly negative for all pairs $j \neq i$. Hence we may define
\begin{align*}
    \gamma \defeq -\max_{(i, j): j \neq i} \left( \frac{\partial f(P,Q)}{\partial p_j} \Bigg|_{P = E_i} - \frac{\partial f(P,Q)}{\partial p_i} \Bigg|_{P = E_i}\right)\\
    \beta \defeq -\min_{(i, j): j \neq i} \left( \frac{\partial f(P,Q)}{\partial p_j} \Bigg|_{P = E_i} - \frac{\partial f(P,Q)}{\partial p_i} \Bigg|_{P = E_i}\right)
\end{align*}
as finite constants, which are strictly positive. As a consequence, we have 
\begin{align}\label{eq: expanding 1st order}
     -\beta \left(1- p_i \right) \leq (P-E_i)^{\intercal} \nabla f(E_i,Q) \leq -\gamma \left(1- p_i \right) 
\end{align}
Once again we use \eqref{eq:diff_grad}, which is 
\begin{align}
    -(P-E_i)^{\intercal} \nabla f(E_i,Q) - \epsilon \left\lVert P-E_i \right\rVert_1 & < f(E_i,Q) -  f(P,Q)\nonumber\\  &\leq -(P-E_i)^{\intercal} \nabla f(E_i,Q) + \epsilon \left\lVert P-E_i \right\rVert_1\nonumber
\end{align}
with $\epsilon= \frac{\gamma}{4}$. By applying \eqref{eq:norm} and the inequality \eqref{eq: expanding 1st order}, we get
\begin{align}
   \frac{\gamma}{2}(1-p_i)  < f(E_i,Q) -  f(P,Q)\leq \left(\beta - \frac{\gamma}{2}\right)(1-p_i),
\end{align}
which holds in the neighbourhoods specified by $\left\lVert P-E_i \right\rVert_1 \leq \delta_{\epsilon}$ of all extreme points $E_i$. This completes the proof of the second statement with $A= \frac{\gamma}{2}>0$, $B= \left(\beta - \frac{\gamma}{2}\right)>0,$ and $a=b=0$.  
}

\bibliographystyle{IEEEtran}
\bibliography{IEEEabrv, main}

\end{document}